\UseRawInputEncoding
\documentclass[a4paper,11pt]{article}
\pdfoutput=1 

\usepackage{jheppub} 

\usepackage[T1]{fontenc} 
\usepackage{amsmath,graphicx,amssymb,subfigure}
\usepackage{slashed}
\usepackage{subfigure}
\usepackage{isomath}
\usepackage{amsthm}
\usepackage{enumerate}
\usepackage{stmaryrd}
\usepackage{amsfonts}
\usepackage{graphics}
\usepackage{epsfig}
\usepackage{slashed}
\usepackage{amsmath}
\usepackage{bm}
\newcommand{\be}{\begin{equation}}
\newcommand{\ee}{\end{equation}}
\newcommand{\bea}{\begin{eqnarray}}
\newcommand{\eea}{\end{eqnarray}}
\newcommand{\myvec}[1]%
{\stackrel{\raisebox{-2pt}[0pt][0pt]{\small$\rightharpoonup$}}{#1}}

\title{\boldmath Pole-skipping and hydrodynamic analysis in Lifshitz, AdS$_2$ and Rindler geometries}

\author[1]{Haiming Yuan}
\author[1,*]{Xian-Hui Ge}

\affiliation[1]{Department of Physics, College of Sciences, Shanghai University, 200444 Shanghai, China}

\emailAdd{gexh@shu.edu.cn}

\abstract{The ``pole-skipping'' phenomenon reflects that the retarded Green's function is not unique at a pole-skipping point in momentum space $(\omega,k)$.   We explore the universality of the pole-skipping in different geometries. In holography, near horizon analysis of the bulk equation of motion is a simpler way to derive a pole-skipping point and we use this method in Lifshitz, AdS$_2$ and Rindler geometries. We also study the complex hydrodynamic analyses and find that the dispersion relations in terms of dimensionless variables $\frac{\omega}{2\pi T}$ and $\frac{\vert k\vert}{2\pi T}$ pass through pole-skipping points $(\frac{\omega_n}{2\pi T}, \frac{\vert k_n\vert}{2\pi T}$) at small $\omega$ and $k$ in Lifshitz background. We verify that the position of the pole-skipping points does not depend on the standard quantization or alternative quantization in the boundary theory in AdS$_2\times\mathbb{R}^{d-1}$ geometry. In Rindler geometry, we cannot find the corresponding Green's function to calculate pole-skipping points because it is difficult to impose the boundary condition. However we can obtain ``special points'' near horizon where bulk equations of motion have two incoming solutions. These ``special points'' correspond to nonunique of the Green's function in physical meaning from the perspective of holography.}

\begin{document}
\maketitle
\flushbottom
\section{Introduction}
\label{sec:intro}
\qquad The holography and the AdS/CFT correspondence \cite{maldacena,witten1,witten2,gubser} provide us a useful method to compute the Green's function at strong coupling and quantum many-body systems \cite{casalderrey,natsuume,ammon,zaanen,hartnoll}. The out-of-time-ordered correlation functions(OTOCs) have been used to investigate the holographic chaos \cite{Shenker1,Roberts,Shenker2}. A new aspect of quantum chaos have been found by using the AdS/CFT correspondence \cite{Makoto1,Makoto2}. We can study the holographic chaos by using the retarded Green's function. The Green's function is not unique at pole-skipping point in complex momentum space $(\omega,k)$ and this phenomenon is known as ``pole-skipping'' \cite{Grozdanov1,Blake,Grozdanov2}. The retarded Green's function is given by
\bea
G^R(\omega,k)_{T^{00}T^{00}}=\frac{b(\omega,k)}{a(\omega,k)}.
\eea
The location of the special points make the coefficient $a(\omega_\star, k_\star)=b(\omega_\star, k_\star)=0$. Then, the retarded Green's function become $G^R(\omega_\star, k_\star)=0/0$. The Green's function depends on the slope $\delta k/\delta \omega$.
\bea
G^R=\frac{(\partial_\omega b)_\star +\frac{\delta k}{\delta \omega}(\partial_k b)_\star+\dots}{(\partial_\omega a)_\star +\frac{\delta k}{\delta \omega}(\partial_k a)_\star+\dots}.
\eea
So if we find the intersection of zeros and poles in the retarded Green's functions, we can obtain these special points. We can use the simpler method, the AdS/CFT duality, to solve special points from bulk field equation \cite{Makoto1,Makoto2,BlakeDavison,Makoto3}. On the bulk field side, there is no unique incoming mode at the horizon and is similar to the ``pole-skipping'' phenomenon in holographic chaos.\\
\indent These special points $(\omega_\star,k_\star)$ can be divided into two classes: one class involves positive imaginary frequencies, while the others are at negative imaginary frequencies. The upper-half $\omega$-plane special point contain the information of quantum chaos. We can extract the Lyapunov exponent $\lambda$ and the butterfly velocity $v_B$ from it
\bea
\label{eq:32}
&&C(t,x)\simeq e^{\lambda(t-x/v_B)}=e^{-i\omega_\star t+ik_\star x},\nonumber\\
&&\omega_\star:=i\lambda,\qquad k_\star:=i\frac{\lambda}{v_B},
\eea
where $C(t,x)$ is a OTOC. $\omega_\star$ and $k_\star$ denote the frequency and momentum at the ``pole-skipping'' point.\\
\indent Although the special points which are located at lower-half $\omega$-plane are not related to the informations of quantum chaos, the retarded Green's functions are also not unique at these special points. The general pole-skipping points in the lower-half $\omega$-plane are located at negative integer (imaginary) Matsubara frequencies $\mathfrak{w}_{n}=-in$ $(n=1,2\dots)$, where $\mathfrak{w}=\frac{\omega}{2\pi T}$. These special points have been found in BTZ black hole \cite{BlakeDavison},  Schwarzschild-AdS spacetime \cite{Makoto3}, 2D CFT \cite{Das}, a holographic system with chiral anomaly \cite{Abbasi1}, a holographic system at finite chemical potential \cite{Abbasi2}, hyperbolic space\cite{Yongjun1}, the large $q$ limit of SYK chain  \cite{Choi} and anisotropic plasma \cite{Karunava}. The lower-half $\omega$-plane pole-skipping points which are at non-integer values of $i\mathfrak{w}$ have be found in 2D CFT \cite{Das} and the large $q$ limit of SYK chain \cite{Choi}. Although the Green's functions at the non-integer values of $i\mathfrak{w}$ pole-skipping points can not be defined uniquely, the incoming boundary condition is unique. The Green's functions at the non-integer values of $i\mathfrak{w}$ pole-skipping points are not related to the incoming solution at the horizon \cite{Yongjun2}. So if we use the holographic near horizon analysis to calculate the lower half-plane pole-skipping points, we just obtain the negative integer points.   \\
\indent The boundary retarded Green's functions $G^R(\omega,k)$ of a conserved $U(1)$ charge current operator $J^\mu$ with $\left\langle J^\mu \right\rangle $= 0. The conserved current $J^\mu$ on the boundary is dual to the bulk field $A_\mu$ in a black hole spacetime. In the linear response regime, we find the retarded Green's function $G^R_{zz}(\omega,k)$ \cite{Iqbal}
\bea
G^R_{zz}(k_\mu)=\frac{\omega^2\sigma}{i\omega-D_ck^2},\qquad k=k_z,
\eea
where $\sigma$ is DC conductivity, $z$ is spatial direction and $D_c$ is diffusion constant of charge. The retarded Green's function has a hydrodynamic pole corresponding to the charge diffusion at very small momentum and frequency
\bea
\label{eq:1}
\omega=-iD_ck^2.
\eea
For the momentum diffusion, the corresponding bulk canonical momenta $T^{\alpha\mu}$ yield the retarded correlator \cite{Iqbal}
\bea
G^R_{\alpha z,\alpha z}=\frac{\omega^2\sigma}{i\omega-D_pk^2},
\eea
where $\alpha$ is any spatial direction $x, y$ and $D_p$ is the diffusion of momentum. The hydrodynamic pole corresponding to the momentum diffusion at very small momentum and frequency is
\bea
\label{eq:2}
\omega=-iD_pk^2.
\eea
In \cite{Grozdanov1}, the upper half-plane pole-skipping point is $\omega_\star=i\lambda_L,\ k_\star=\sqrt{6}i\pi T$.
The dispersion relation of hydrodynamic sound modes $\mathfrak{w}=v_B \mathfrak{k}$ passes through the upper half-plane pole-skipping point $(\mathfrak{w}_\star,\mathfrak{k}_\star)$ for the case of Schwarzschild-${\rm AdS_5}$ spacetime, where $\mathfrak{w}=\frac{\omega}{2\pi T}$ and $\mathfrak{k}=\frac{k}{2\pi T}$. In \cite{BlakeDavison}, it was shown that the hydrodynamic poles \eqref{eq:1} and \eqref{eq:2} pass through the pole-skipping points in the lower half-plane for the case of Schwarzschild-AdS spacetime. Ref.\cite{Karunava} shows that the dispersion relation which arises from the pole of the retarded Green's function associated to the transverse momentum density passes through the lower half-plane pole-skipping points in the backgound of anisotropic plasma. The pole-skipping points may be related to the hydrodynamic dispersion relation.\\
\indent In this paper, we show that near horizon analysis not only applies to AdS spacetime, but also to Lifshitz and Rindler geometry. The pole-skipping points are related to the hydrodynamic dispersion relations in Lifshitz geometry. In section~\ref{sec:Aniso}, we calculate the lower half-plane pole-skipping points of tensor , Maxwell vector and  Maxwell scalar mode in the background of the anisotropic system near Lifshitz points. The frequencies of these special points are located at negative integer (imaginary) Matsubara frequencies  $\mathfrak{w}_{n}:=\frac{\omega_n}{2\pi T}=-in$ $(n=1,2,3\dots)$. We study the complex hydrodynamic analyses and the analytic properties of the dispersion relations have been treated as Puiseux series in complex momentum \cite{Grozdanov3}. We plot the hydrodynamic dispersion relations in terms of dimensionless variables $\frac{\omega}{2\pi T}$ and $\frac{\vert k\vert}{2\pi T}$. We find that the hydrodynamic dispersion relations pass through pole-skipping points $(\mathfrak{w}_{n}, \vert\mathfrak{k}_{n}\vert)$ along the direction of translation invariance at very small momentum and frequency. In section~\ref{sec:Lifshitz}, we consider the axion field in the Lifshitz black hole with hyperscaling violating factor. The locations of lower half-plane pole-skipping points are the same at negative integer (imaginary) Matsubara frequencies. The hydrodynamic dispersion relation also fits the pole-skipping points well in the small $k$ limit in this section. In section~\ref{sec:AdS}, we calculate the pole-skipping points of scalar field in the background of AdS$_2\times\mathbb{R}^{d-1}$. The results from the near horizon analysis of the bulk equation are the same as the outcomes from the Green' function. We all know that the Green's function depends on the boundary conditions. For instance, choosing standard quantisation makes the Green's function different from choosing alternative quantisation for a scalar. But for the pole-skipping points, choosing different quantization will not affect their positions\cite{Yongjun1}. We also verify the correctness of this conclusion in this section. In section~\ref{sec:Rindler}, we obtain the ``special points'' of sound mode and scalar field in Rindler horizon. There is no boundary condition in Rindler geometry, so it still not enough to ensure a ``pole-skipping'' in the corresponding Green's function. But it is interesting that we can obtain ``special points'' from near horizon analysis which means there are two incoming solutions near the horizon which are related to the nonunique of the Green's function from a holographic point of view.
\section{Anisotropic system near Lifshitz points}
\label{sec:Aniso}
\qquad In this section, we consider the anisotropic spacetime near Lifshitz points and compute the lower half-plane pole-skipping points in the viscosity tensor modes, Maxwell vector and  Maxwell scalar mode. The form of action is Einstein-Maxwell-Dilaton action given as \cite{Inkof}
\bea
S=\int d^{3+1}x\sqrt{-g}(R+\mathcal{L}_M).
\eea
The matter Lagrangian is given by
\bea
\label{eq:14}
\mathcal{L}_M=-\frac{1}{2}(\nabla\varphi)^2-V(\varphi)-\frac{Y(\varphi)}{2}(\nabla\psi)^2-\frac{Z(\varphi)}{4}F^2.
\eea
$V(\varphi)$, $Y(\varphi)$ and $Z(\varphi)$ are the scalar potentials that take the following form in the IR ($r\rightarrow 0$)
\bea
V_{IR}=-V_0e^{\delta \varphi},\quad Y_{IR}=e^{\lambda \varphi},\quad Z_{IR}=e^{\zeta \varphi}\nonumber
\eea
with the dilaton $\varphi=2\kappa {\rm log}(r)$. The matter Lagrangian become
\bea
\mathcal{L}_M=-\frac{1}{2}(\nabla\varphi)^2+V_0r^{2\kappa\delta}-\sum^p_{\alpha=1}\frac{r^{2\kappa\lambda_\alpha}}{2}(\nabla\psi_\alpha)^2-\frac{r^{2\kappa\zeta}}{4}F^2.
\eea
Where $p$ is the number of axions and $\psi_\alpha=a_\alpha x_\alpha$. The notation $a_\alpha$ is a constant which is a measure of the anisotropy along the direction of $\alpha\,(\alpha=x,y)$. The critical scaling of the near horizon region (IR) is holographically realized by a Lifshitz geometry of the form
\bea
\label{eq:3}
ds^2=r^\theta\bigg(-\frac{f(r)dt^2}{r^{2z}}+L^2\frac{dr^2}{f(r)r^2}+\frac{dx^2}{r^{2\phi}}+\frac{dy^2}{r^2}\bigg),
\eea
where $f(r)=1-(\frac{r}{r_+})^{\delta_0}$, $\delta_0=1+\phi+z-\theta$. $z$ is the Lifshitz scaling exponent, $\theta$ is the hyperscaling violating exponent.
The anisotropy is expressed in terms of the exponent $\phi$ that relates momenta between two directions
\bea
\vert k_x\vert\sim\vert k_y\vert^{\phi}.
\eea
 The relationship between characteristic energies and the momenta in two spatial dimensions
\bea
\omega&\sim&\vert k_x\vert^{z/\phi},\nonumber\\
\omega&\sim&\vert k_y\vert^{z}.
\eea
The Hawking temperature is given by $T=\frac{\vert\delta_0\vert r_+^{-z}}{4\pi L}$. The authors in \cite{Inkof} derived the hyperscaling-violating solutions in the presence of two axion ($p=1$) fields in the IR geometry
\bea
\label{eq:29}
&&z=\phi,\quad 2\kappa\delta=-\theta, \quad \kappa\lambda=-1,\nonumber\\
&&4\kappa^2=\theta^2-2\theta\phi+2\phi-2,\nonumber\\
&&L^2=(\theta-2\phi-1)(\theta-2\phi)/V_0,\nonumber\\
&&a^2=\frac{2V_0(1-\phi)}{\theta-2\phi}.
\eea
Using the Eddington-Finkelstein (EF) coordinates, we recast the tortoise coordinate $dr_{\ast}=\frac{r^{z-1}}{f(r)}dr$ and $v=t-r_{\ast}$ into the metric \eqref{eq:3} and obtain
\bea
ds^2=-r^{\theta-2z}f(r)dv^2+2Lr^{\theta-z-1}dvdr+r^{\theta-2\phi}dx^2+r^{\theta-2}dy^2.
\eea
\subsection{Tensor-type perturbations}
\subsubsection*{$\bullet$ Case one: $x$-direction ($a$=0)}
Firstly, we consider the tensor type of perturbation along the $x$-direction of the form  $\delta h_{xy}=e^{-i\omega v+ik_xx}h_{xy}(r)$. Note that $h_{xy}=g_{xx}\delta h^x_y$. We choose $z=\phi=1$ to make $a=0$, the equation of motion for $h^x_{y}$ is given by \cite{Inkof}
\bea
\label{eq:11}
\partial_\mu(\frac{\sqrt{-g}}{\mathcal{N}}\partial^\mu h^x_y)=0.
\eea
The notations of $\mathcal{N}$ is expressed as
\bea
\mathcal{N}(r)=g_{yy}(r)g^{xx}(r).
\eea
The equation of perturbation become
\bea
\label{eq:4}
&&h''^x_y+\bigg(\frac{f'(r)}{f(r)}-\frac{2i\omega L}{f(r)}+\frac{(\theta-2)}{r}\bigg)h'^x_y-\frac{L}{f(r)}\bigg(k^2_x L+i\omega r^{-1}(\theta-2)\bigg)h^x_y=0.
\eea
We use the approximation $f(r)\sim f'(r_+)(r-r_+)$ near the horizon $r=r_+$, then expand the field equation near the horizon $r=r_+$
\bea
h''^x_y+\frac{1-i\mathfrak{w}}{r-r_{+}}h'^x_y+\frac{L}{\delta_0}\bigg(4\mathfrak{k}_x^{2}\pi^2T^2Lr_++i2\pi T\mathfrak{w} (\theta-2)\bigg)\frac{h^x_y}{r-r_{+}}=0,\nonumber\\
\eea
where $\mathfrak{w}=\frac{\omega}{2\pi T}$, and $\mathfrak{k}_x=\frac{k_x}{2\pi T}$. For a generic $(\mathfrak{w},\mathfrak{k}_x)$, the equation has a regular singularity at $r=r_+$. So one can solve it by a power series expansion around $r=r_+$
\bea
\label{eq:5}
h^y_x(r)=(r-r_+)^\lambda \sum^\infty_{p=0}h^y_{xp}(r-r_+)^p.
\eea
At the lowest order, we can obtain the indicial equation $\lambda(\lambda-i\mathfrak{w})=0$ and two solutions
\bea
\label{eq:6}
\lambda_1=0,\qquad \lambda_2=i\mathfrak{w}.
\eea
One denotes the incoming mode and another the outgoing mode. If we choose $i\mathfrak{w}=1$ and appropriately $\mathfrak{k}_x$ make the singularity in front of $h'^y_x$ and $h^y_x$ terms vanishing. We call it a ``pole-skipping''. The regular singularity at $r=r_+$ becomes a regular point at this special point. We make the coefficients $(1-i\mathfrak{w})$ and $\bigg(4\mathfrak{k}_x^{2}\pi^2T^2Lr_++i2\pi T\mathfrak{w} (\theta-2)\bigg)$ become 0. We then obtain the location of the special point about $h^y_x$ field equation
\bea
\label{eq:7}
&&\mathfrak{w}_{\star1}=-i,\nonumber\\
&&\mathfrak{k}^{2}_{x\star1}=\frac{2(2-\theta)}{\vert\delta_0\vert}.
\eea
When $\theta=0$ and $L^2=1$, the metric \eqref{eq:3} recovers the black brane solution in $AdS_4$ and the pole-skipping point become
\bea
\mathfrak{w}_{\star}=-i,\quad \mathfrak{k}^{2}_{x\star}=\frac{4}{3}.
\eea
This result is the same as (1.7e) in ref.\cite{Makoto2}. \\
\indent From \eqref{eq:7}, two solutions in \eqref{eq:6} become a regular solution
\bea
\lambda_1=0,\qquad \lambda_2=1.
\eea
So that \eqref{eq:5} can be written as Taylor series. The special point above is $(\omega_1=-i2\pi T, k_1)$. We extend pole-skipping phenomenon at higher Matsubara frequencies $\omega_n=-i2\pi Tn$, and the method has been showed in \cite{BlakeDavison}. Now we review this process calculating special points $(\omega_n, k_n)$. Firstly we expand $h^y_x(r)$ with a Taylor series
\bea
\label{eq:8}
h^y_x(r)=\sum^\infty_{p=0}h^y_{xp}(r-r_+)^p=h^y_{x0}+h^y_{x1}(r-r_+)+h^y_{x2}(r-r_+)^2+\dots.
\eea
We insert \eqref{eq:8} into \eqref{eq:4} and expand the axion equation of motion in powers of $(r-r_+)$. Then a series of perturbed equation in the order of $(r-r_+)$ can be denoted as
\bea
\label{eq:9}
S=\sum^\infty_{p=0}S_p(r-r_+)^p=S_0+S_1(r-r_+)+S_2(r-r_+)^2+\dots.
\eea
We write down the first few equation $S_p=0$ in the expansion of \eqref{eq:9}
\bea
&&0=M_{11}(\omega,k^2)h^y_{x0}+(2\pi T-i\omega)h^y_{x1},\nonumber\\
&&0=M_{21}(\omega,k^2)h^y_{x0}+M_{22}(\omega,k^2)h^y_{x1}+(4\pi T-i\omega)h^y_{x2},\nonumber\\
&&0=M_{31}(\omega,k^2)h^y_{x0}+M_{32}(\omega,k^2)h^y_{x1}+M_{33}(\omega,k^2)h^y_{x2}+(6\pi T-i\omega)h^y_{x3},
\eea
In order to find an incoming solution, we should solve a set of linear equations of the form
\bea
\label{eq:25}
\mathcal{M}^{(n)}(\omega,k^2)\cdot h^y_{x}\equiv\left(\begin{array}{ccccc}
    M_{11} & (2\pi T-i\omega) & 0    & 0  &\dots\\
    M_{21} & M_{22}& (4\pi T-i\omega)& 0   &\dots\\
    M_{31} & M_{32}&  M_{33} &(6\pi T-i\omega) &\dots\\
    \dots   &  \dots&  \dots  &\dots   &\dots\\
\end{array}\right)\left(\begin{array}{ccccc}
    h^y_{x0}\\
   h^y_{x1}\\
    h^y_{x2} \\
    \dots \\
\end{array}\right)=0
\eea
\begin{figure}[!t]
\begin{minipage}{0.48\linewidth}
\centerline{\includegraphics[width=7.5cm]{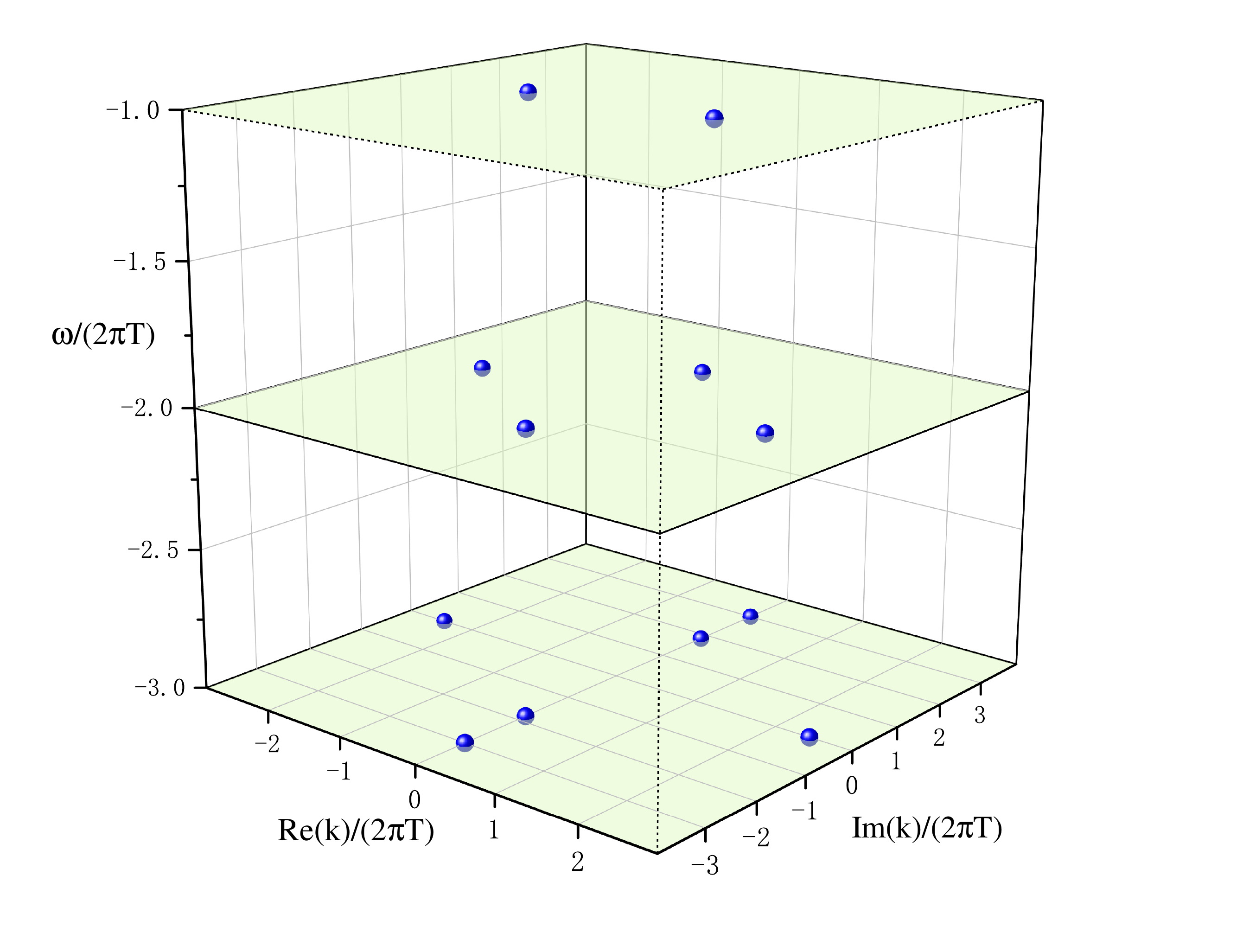}}
\centerline{(a)}
\end{minipage}
\hfill
\begin{minipage}{0.48\linewidth}
\centerline{\includegraphics[width=7.5cm]{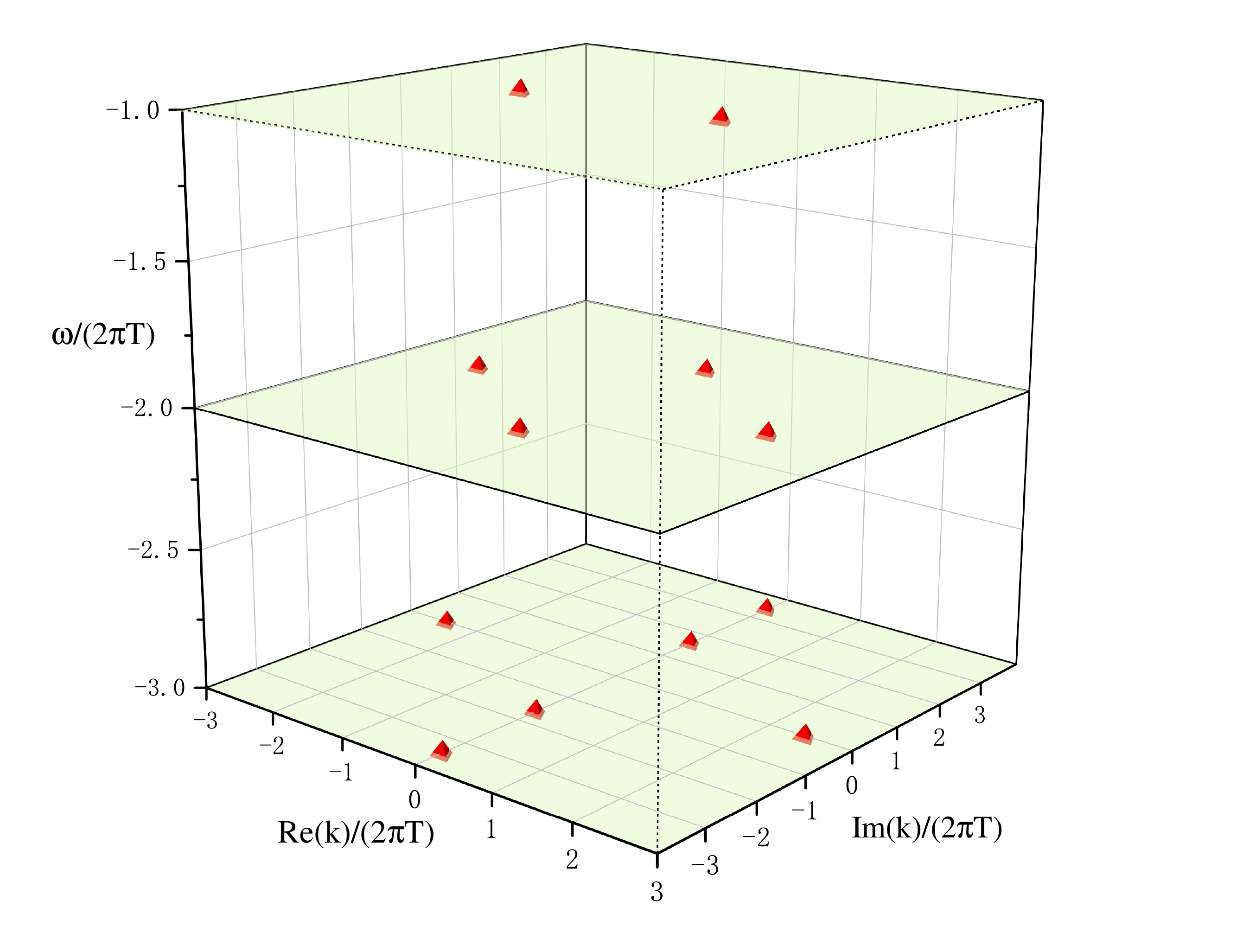}}
\centerline{(b)}
\end{minipage}
\vfill
\caption{\label{fig:Figure1}The first three order pole-skipping points of tensor mode $h^y_x$ along the $x$-direction in anisotropic system near Lifshitz points when $a=0$. (a) $\theta=0$, $z=\phi=1$; (b) $\theta=-3$, $z=\phi=1$. }
\end{figure}
The locations of special points $(\omega_n, k_n)$  can be easily extracted from determinant of the $(n\times n)$ matrix $\mathcal{M}^{(n)}(\omega,k^2)$ constructed by the first $n$ equations
\bea
\omega_n=-i2\pi Tn,\qquad k^2=k^2_n,\qquad {\rm det}\mathcal{M}^{(n)}(\omega,k^2)=0.
\eea
\bea
&&\mathfrak{w}_{\star2}=-2i,\nonumber\\
&&\mathfrak{k}^{2}_{x\star2,1}=\frac{4\pi TL(2-\theta)+r_+f''(r_+)-\sqrt{16\pi^2T^2L^2(12-8\theta+\theta^2)+8\pi TLr_+(\theta-2)f''(r_+)+r_+^2f''^2(r_+)}}{8\pi^2T^2L^2r_+},\nonumber\\
&&\mathfrak{k}^{2}_{x\star2,2}=\frac{4\pi TL(2-\theta)+r_+f''(r_+)+\sqrt{16\pi^2T^2L^2(12-8\theta+\theta^2)+8\pi TLr_+(\theta-2)f''(r_+)+r_+^2f''^2(r_+)}}{8\pi^2T^2L^2r_+}.\nonumber\\
\eea
\begin{figure}[htp]
	\begin{centering}
		\includegraphics[scale=1]{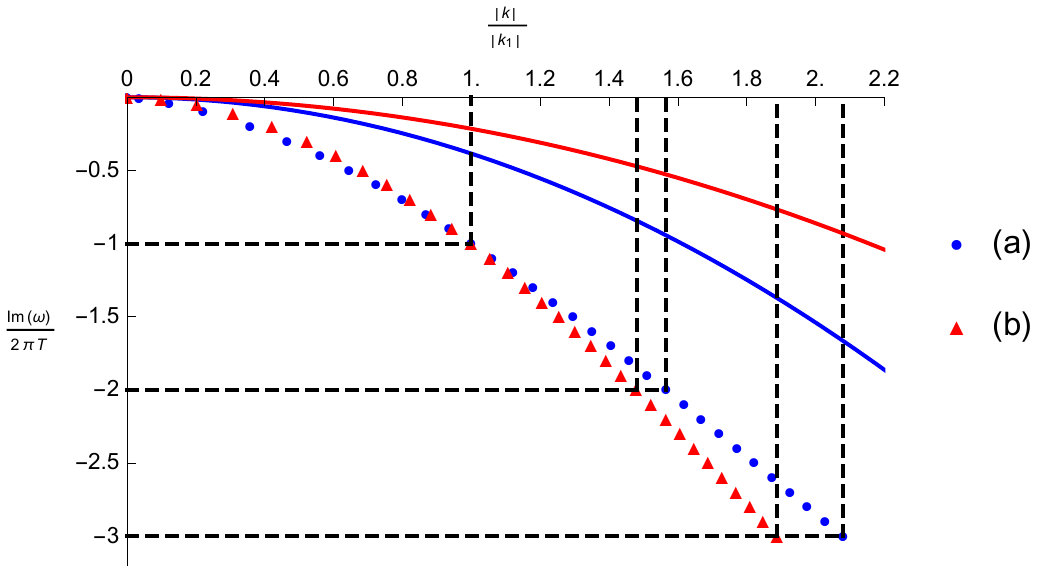}
		\par\end{centering}
	\caption{\label{fig:Figure2} Solid lines show the diffusive hydrodynamic dispersion relation\eqref{eq:10}. The intersections of the dashed lines correspond to the first three order pole-skipping points of tensor mode $h^y_x$ along the $x$-direction in the anisotropic system near Lifshitz points when $a=0$. Blue dots and red triangles show the curve fitting of pole-skipping points of the data which we set (a) $z=\phi=1$, $\theta=0$ ; (b) $z=\phi=1$, $\theta=-3$. }
\end{figure}
At these locations, any value of $h^y_{x0}$ and $h^y_{xn}$ satisfies the equation \eqref{eq:4} and thus there are two independent free parameters $h^y_{x0}$ and $h^y_{xn}$  in the general series solution \eqref{eq:8} to this equation. The general ingoing solution to the equation of motion \eqref{eq:4} is not unique at Matsubara frequencies $\omega_n=-i2\pi Tn$. The first few elements of this matrix have showed in appendix A.1.1. Then we calculate the first three order pole-skipping points and plot them in figure~\ref{fig:Figure1}.\\
\textbf{(a)} $\theta=0$, $\phi=z=1$
\bea
\mathfrak{w}_{\star1}&=&-i,\; \quad \mathfrak{k}_{x\star1}=\pm1.155;\nonumber\\
\mathfrak{w}_{\star2}&=&-2i,\quad \mathfrak{k}_{x\star2,1}=\pm1.807,\ \mathfrak{k}_{x\star2,2}=\pm1.807i;\nonumber\\
\mathfrak{w}_{\star3}&=&-3i,\quad \mathfrak{k}_{x\star3,1}=\pm2.402,\ \mathfrak{k}_{x\star3,2}=\pm1.849i,\ \mathfrak{k}_{x\star3,3}=\pm3.002i.
\eea
\textbf{(b)} $\theta=-3$, $\phi=z=1$
\bea
\mathfrak{w}_{\star1}&=&-i,\; \quad \mathfrak{k}_{x\star1}=\pm1.291;\nonumber\\
\mathfrak{w}_{\star2}&=&-2i,\quad \mathfrak{k}_{x\star2,1}=\pm1.911,\ \mathfrak{k}_{x\star2,2}=\pm1.911i;\nonumber\\
\mathfrak{w}_{\star3}&=&-3i,\quad \mathfrak{k}_{x\star3,1}=\pm1.633i,\ \mathfrak{k}_{x\star3,2}=\pm2.438,\ \mathfrak{k}_{x\star3,3}=\pm3.407i.
\eea
\indent The transverse momentum Green's function has a hydrodynamic pole corresponding to the diffusion of momentum with the small $k$ dispersion relation \cite{Iqbal}
\bea
\label{eq:10}
\omega(k)=-iD_pk^2+\dots,
\eea
Because $a=0$, the momentum along $x$-direction is without dissipation. So $\frac{\eta_{yxyx}}{s}=\frac{1}{4\pi}\frac{g_{xx}}{g_{yy}}\vert_{r_+}$ \cite{Inkof}. We obtain the momentum diffusion constant $D_p=\eta/(sT)=1/(4\pi T)$. Because of the complex momentum $k_n$,  we study the complex hydrodynamic analyse.  We plot first three order pole-skipping points  of $(\mathfrak{w}_{n}, \vert\mathfrak{k}_{n}\vert)$  and hydrodynamic dispersion relation in terms of dimensionless variables $\frac{\omega}{2\pi T}$ and $\frac{\vert k\vert}{2\pi T}$  in figure~\ref{fig:Figure2}. We can see that the dispersion relation $\omega(k)$ of a hydrodynamic mode passes through pole-skipping points when $\omega, k\rightarrow 0$.
\subsubsection*{$\bullet$ Case two: $x$-direction ($a$ $\neq$ 0)}
\indent Now we choose $z=\phi\neq 1$ to make $a\neq 0$. The equation of motion for $h^x_{y}$ is given as \cite{Inkof}
\bea
\label{eq:12}
\partial_\mu(\frac{\sqrt{-g}}{\mathcal{N}}\partial^\mu h^x_y)-\frac{\sqrt{-g}}{\mathcal{N}}m^2h^x_y=0.
\eea
The notation of $m^2$ is expressed as $m^2(r)=a^2Y(\varphi)g^{xx}(r)$. In this case, translational symmetry along the $x$-direction is broken and momentum is dissipated at a strength controlled by $a$.  If we drop the mass term  $m^2$, the equation is the same as the isotropic one \eqref{eq:11} discussed in the previous case.\\
\indent The equation \eqref{eq:12} can be recast as
\bea
\label{eq:26}
&&h''^x_y+\bigg(\frac{f'(r)}{f(r)}-\frac{2i\omega Lr^{\phi-1}}{f(r)}+(\theta-4\phi+2)r^{-1}\bigg)h'^x_y\nonumber\\
&&-\frac{L}{f(r)}\bigg(a^2 Lr^{2(\kappa\lambda+\phi)-2}+k_x^2Lr^{2\phi-2}+i\omega r^{\phi-2}(\theta-3\phi+1)\bigg)h^x_y=0.
\eea
We also use the approximation $f(r)\sim f'(r_+)(r-r_+)$ near the horizon $r=r_+$. We expand the field equation near the horizon $r=r_+$
\bea
h''^y_x+\frac{1-i\mathfrak{w}}{r-r_{+}}h'^x_y+\frac{L}{\delta_0}\bigg(a^2 Lr^{2(\kappa\lambda+\phi)-1}+4\mathfrak{k}_x^{2}\pi^2T^2Lr^{2\phi-1}+i2\pi T\mathfrak{w} r^{\phi-1}(\theta-3\phi+1)\bigg)\frac{h^x_y}{r-r_{+}}=0,\nonumber\\
\eea
We take the value of coefficients $(1-i\mathfrak{w})$ and $\frac{L}{\delta_0}\bigg(a^2 Lr^{2(\kappa\lambda+\phi)-1}+4\mathfrak{k}_x^{2}\pi^2T^2Lr^{2\phi-1}+i2\pi T\mathfrak{w} r^{\phi-1}(\theta-3\phi+1)\bigg)$ become 0 to eliminate the singularity in front of $h'^y_x$ and $h^y_x$ terms, then we find the location of the special point
\bea
&&\mathfrak{w}_{\star1}=-i,\nonumber\\
&&\mathfrak{k}^{2}_{x\star1}=\frac{-2\vert\delta_0\vert(\theta-3\phi+1)-4a^2L^2r_+^{2(\kappa\lambda+\phi)}}{\vert\delta_0\vert^2}.
\eea
If we set $\phi=1$ to make $a=0$, we can see this result becomes \eqref{eq:7}. We can also evaluate the higher special points
\bea
\omega_n=-i2\pi Tn,\qquad k^2=k^2_n,\qquad {\rm det}\mathcal{M}^{(n)}(\omega,k^2)=0.
\eea
\bea
\mathfrak{w}_{\star2}&=&-2i,\nonumber\\
\mathfrak{k}^{2}_{x\star2,1}&=&\frac{r_+^{-2\phi}}{8\pi^2T^2L^2}\bigg(r_+^2f''(r_+)-2a^2L^2r_+^{2(\kappa\lambda+\phi)}-4\pi TLr_+^\phi(\theta-2\phi)-\sqrt{16\pi TL^2r_+^{2\phi}(2a^2Lr_+^{2\kappa\lambda+\phi}\kappa\lambda} \nonumber\\
&&\overline{+\pi T(4+4\theta+\theta^2-20\phi-12\theta\phi+28\phi^2))-8\pi TLr_+^{\phi+2}(4\phi-\theta-2)f''(r_+)+r_+^4f''^2(r_+)}\bigg),\nonumber\\
\mathfrak{k}^{2}_{x\star2,2}&=&\frac{r_+^{-2\phi}}{8\pi^2T^2L^2}\bigg(r_+^2f''(r_+)-2a^2L^2r_+^{2(\kappa\lambda+\phi)}-4\pi TLr_+^\phi(\theta-2\phi)+\sqrt{16\pi TL^2r_+^{2\phi}(2a^2Lr_+^{2\kappa\lambda+\phi}\kappa\lambda} \nonumber\\
&&\overline{+\pi T(4+4\theta+\theta^2-20\phi-12\theta\phi+28\phi^2))-8\pi TLr_+^{\phi+2}(4\phi-\theta-2)f''(r_+)+r_+^4f''^2(r_+)}\bigg).\nonumber\\
\eea
\begin{figure}[!t]
\begin{minipage}{0.48\linewidth}
\centerline{\includegraphics[width=7.5cm]{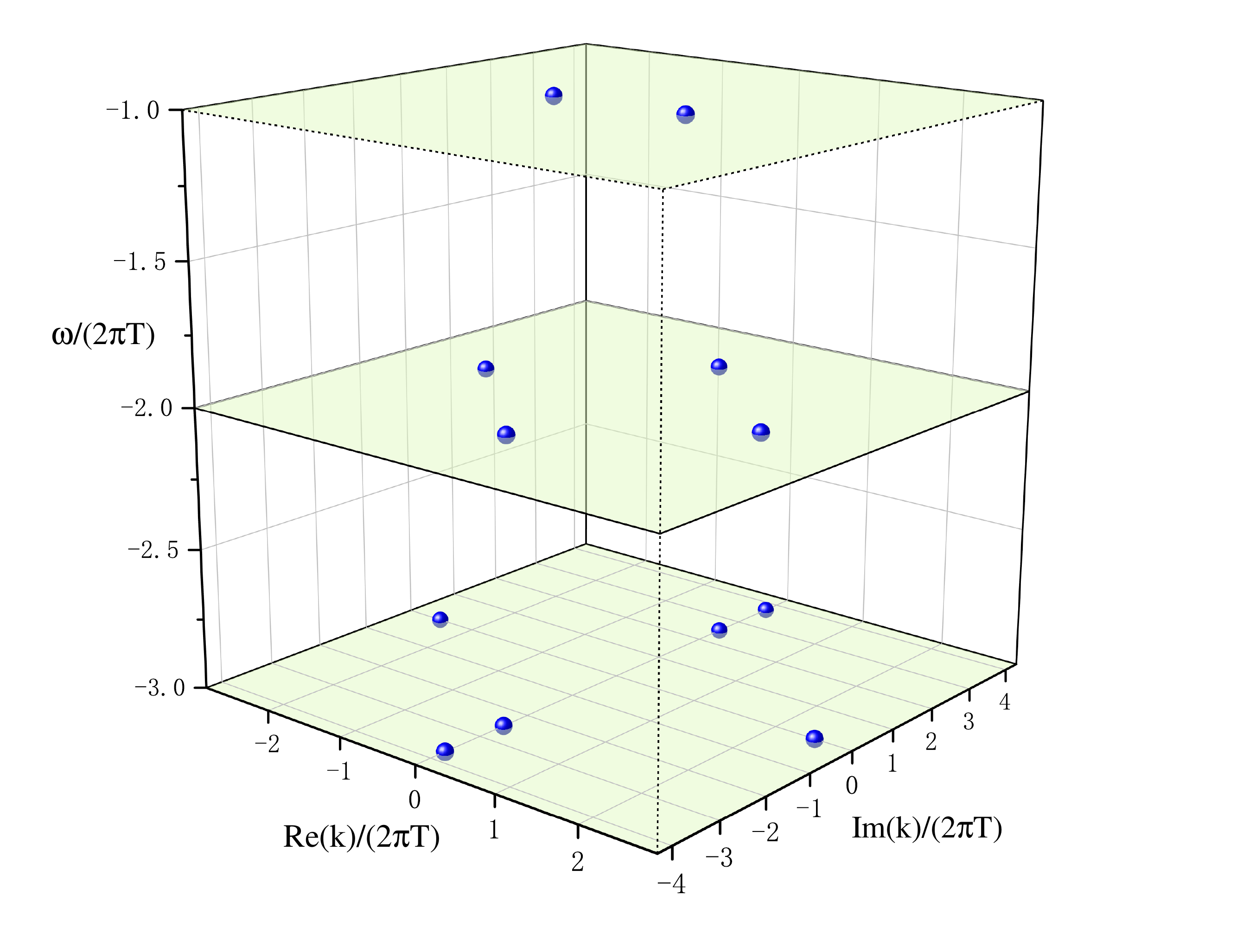}}
\centerline{(a)}
\end{minipage}
\hfill
\begin{minipage}{0.48\linewidth}
\centerline{\includegraphics[width=7.5cm]{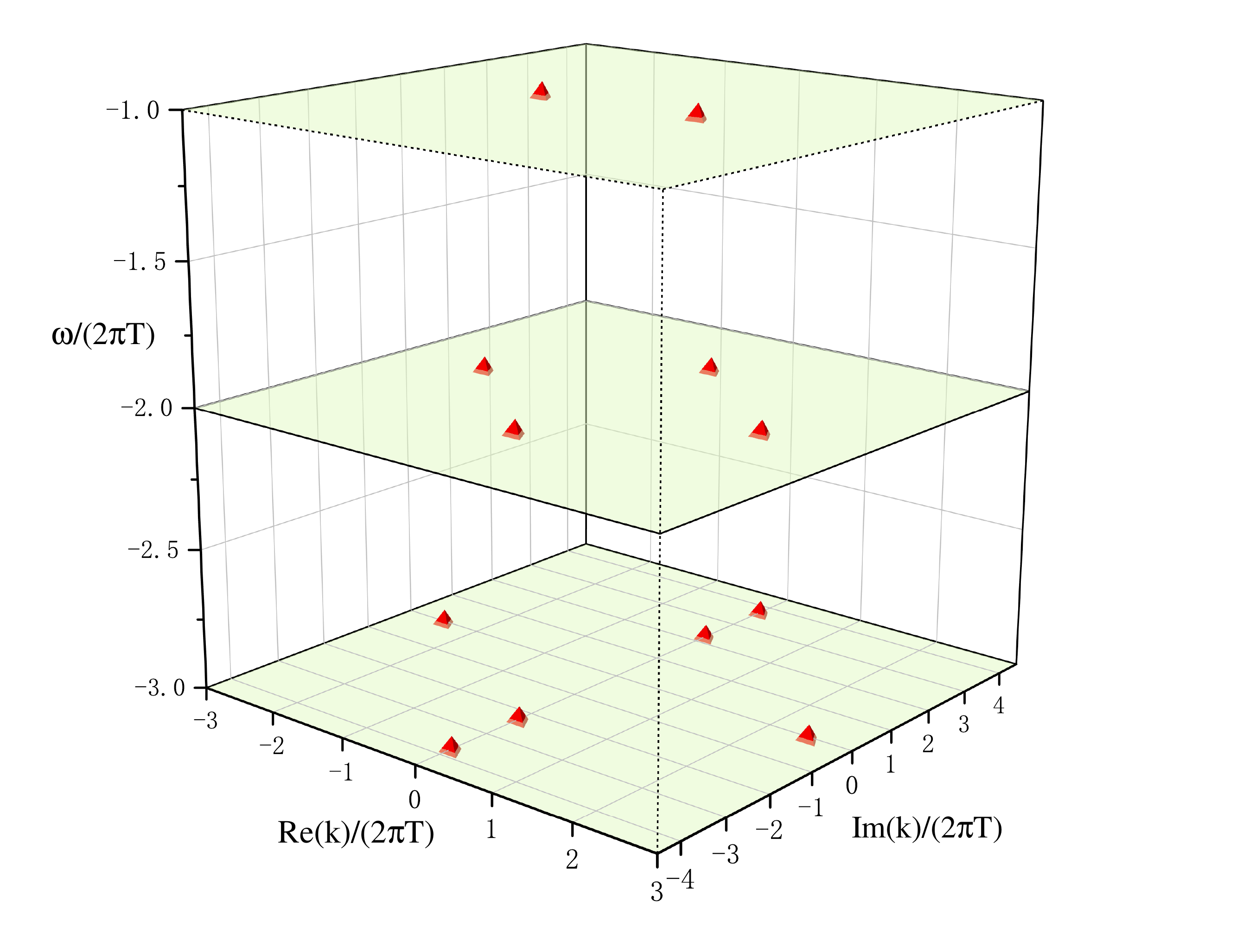}}
\centerline{(b)}
\end{minipage}
\vfill
\caption{\label{fig:Figure3} The first three order pole-skipping points of tensor mode $h^y_x$ along the $x$-direction in the anisotropic system near Lifshitz points when $a\neq 0$. (a) $\theta=-1$, $z=\phi=2$ ($a=3.46$); (b) $\theta=-3$, $z=\phi=2$ ($a=4$). }
\end{figure}
The first few elements of this matrix have showed in appendix A.1.2. Then we calculate the first three order pole-skipping points and plot them in the figure~\ref{fig:Figure3}. We choose the following parameters \\
\textbf{(a)} $\theta=-1$, $z=\phi=2$ ($a=3.46$)
\bea
\mathfrak{w}_{\star1}&=&-i,\; \quad \mathfrak{k}_{x\star1}=\pm0.816;\nonumber\\
\mathfrak{w}_{\star2}&=&-2i,\quad \mathfrak{k}_{x\star2,1}=\pm1.758,\ \mathfrak{k}_{x\star2,2}=\pm2.400i;\nonumber\\
\mathfrak{w}_{\star3}&=&-3i,\quad \mathfrak{k}_{x\star3,1}=\pm2.462,\ \mathfrak{k}_{x\star3,2}=\pm2.506i,\ \mathfrak{k}_{x\star3,3}=\pm3.713i.
\eea
\textbf{(b)} $\theta=-3$, $\phi=z=2$ ($a=4$)
\bea
\mathfrak{w}_{\star1}&=&-i,\; \quad \mathfrak{k}_{x\star1}=\pm1;\nonumber\\
\mathfrak{w}_{\star2}&=&-2i,\quad \mathfrak{k}_{x\star2,1}=\pm1.833,\ \mathfrak{k}_{x\star2,2}=\pm2.345i;\nonumber\\
\mathfrak{w}_{\star3}&=&-3i,\quad \mathfrak{k}_{x\star3,1}=\pm2.270i,\ \mathfrak{k}_{x\star3,2}=\pm2.478,\ \mathfrak{k}_{x\star3,3}=\pm3.741i.
\eea
\subsection{Sound modes}
The sound modes of metric perturbation are given as
\bea
\label{eq:13}
h_{vv}=e^{-i\omega v+ik_xx}h_{vv}(r),\quad h_{vx}=e^{-i\omega v+ik_xx}h_{vx}(r),\nonumber\\
h_{xx}=e^{-i\omega v+ik_xx}h_{xx}(r),\quad h_{yy}=e^{-i\omega v+ik_xx}h_{yy}(r).
\eea
Substituting \eqref{eq:13} into the linearized Einstein equation
\bea
R_{\mu\nu}-\frac{1}{2}R g_{\mu\nu}=-\frac{1}{\sqrt{-g}}\frac{\delta(\sqrt{-g}\mathcal{L}_M)}{\delta g^{\mu\nu}}.
\eea
Where $\mathcal{L}_M$ is given by eq.\eqref{eq:14}. We obtain the $vv$ component near the horizon $r_+$
\bea
L(2kr_+^{2\phi}h_{tx}+r_+^{2\phi}\omega h_{xx}+r_+^{2}\omega h_{yy})(\omega-2i\pi T)+h_{tt}(Lk_x^2r_+^{2\phi}-Lr_+^\theta \mathcal{L}_M+r_+^\phi(\phi-\theta+1)(i\omega-4\pi T)).\nonumber\\
\eea
The lower-half $\omega$-plane of pole-skipping point is located at
\bea
\omega_{\star}=i2\pi T,\quad k_{x\star}^{2}=\frac{6\pi T(\phi-\theta+1)r_+^{-\phi}}{L}+\mathcal{L}_M  r_+^{\theta-2\phi}.
\eea
The Lyapunov exponent and butterfly velocity can be calculated by eq.\eqref{eq:32}
\bea
\lambda_L=2\pi T,\quad v_B^2=\frac{\vert\omega\vert^2}{\vert k_x^2\vert}=\frac{4\pi^2T^2L^2r^{4\phi}}{(L\mathcal{L}_Mr_+^\theta+6\pi T (\phi-\theta+1)r_+^\phi)^2}.
\eea
\subsection{Maxwell field}
\subsubsection{Maxwell vector mode}
\qquad We consider the Maxwell vector perturbation $A_y=a_ye^{-i\omega v+ik_x x}$ in the probe limit. The Maxwell vector equation is written as \cite{Inkof}
\bea
\label{eq:15}
\partial_\nu(\sqrt{-g}Z(\varphi)F^{\mu\nu})=0
\eea
In the region of IR , $Z(\varphi)=r^{2\kappa\zeta}$. Substituting it into \eqref{eq:15}, the equation become
\bea
\label{eq:31}
&&A''_y+\bigg(\frac{f'(r)}{f(r)}-\frac{2i\omega r^{z-1}}{f(r)}+(2-z+2\kappa\zeta-\phi)r^{-1}\bigg)A'_y\nonumber\\
&&-\frac{1}{f(r)}\bigg(k^2_x r^{2\phi-2}+i\omega(1+2\kappa\zeta-\phi)r^{z-2}\bigg)A_y=0.
\eea
Using the approximation $f(r)\sim f'(r_+)(r-r_+)$ near the horizon $r=r_+$, we expand the field equation near the horizon $r=r_+$
\bea
A''_y+\frac{1-i\mathfrak{w}}{r-r_{+}}A'_y-\frac{1}{\vert\delta_0\vert}\bigg(4\mathfrak{k}^{2}_{x}\pi^2T^2r^{2\phi-2}+i2\pi T\mathfrak{w}(1+2\kappa\zeta-\phi)r^{z-2}\bigg)\bigg|_{r=r_+}\frac{A_y}{r-r_+}=0.\nonumber\\
\eea
We take the value of coefficients $(1-i\mathfrak{w})$ and $\frac{1}{\vert\delta_0\vert}\bigg(4\mathfrak{k}^{2}_{x}\pi^2T^2r^{2\phi-2}+i2\pi T\mathfrak{w}(1+2\kappa\zeta-\phi)r^{z-2}\bigg)\bigg|_{r=r_+}$ to be 0 and eliminate the singularity in front of $A'_y$ and $A_y$ terms. The location of the special point about Maxwell vector mode is
\begin{figure}[!t]
\begin{minipage}{0.48\linewidth}
\centerline{\includegraphics[width=7.5cm]{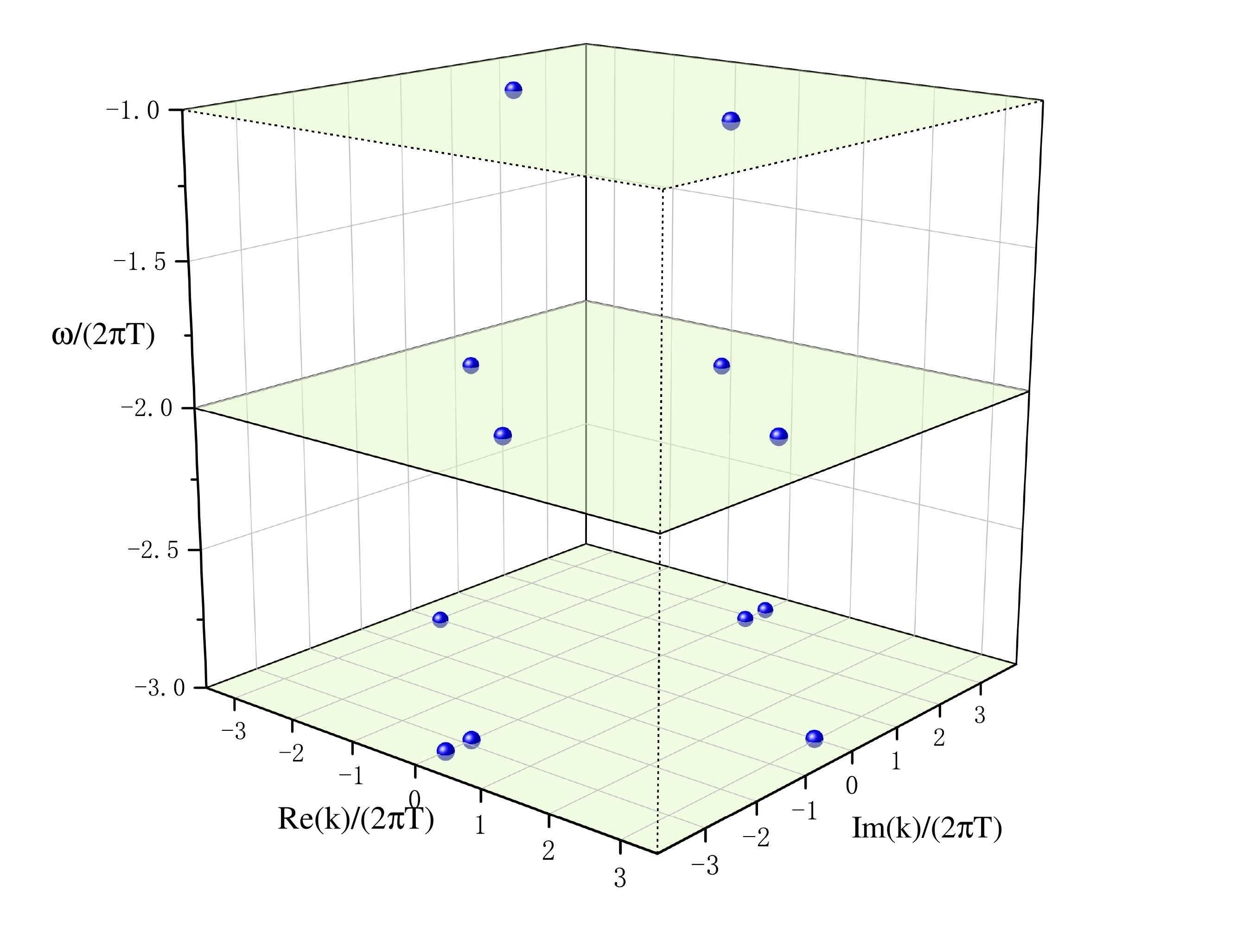}}
\centerline{(a)}
\end{minipage}
\hfill
\begin{minipage}{0.48\linewidth}
\centerline{\includegraphics[width=7.5cm]{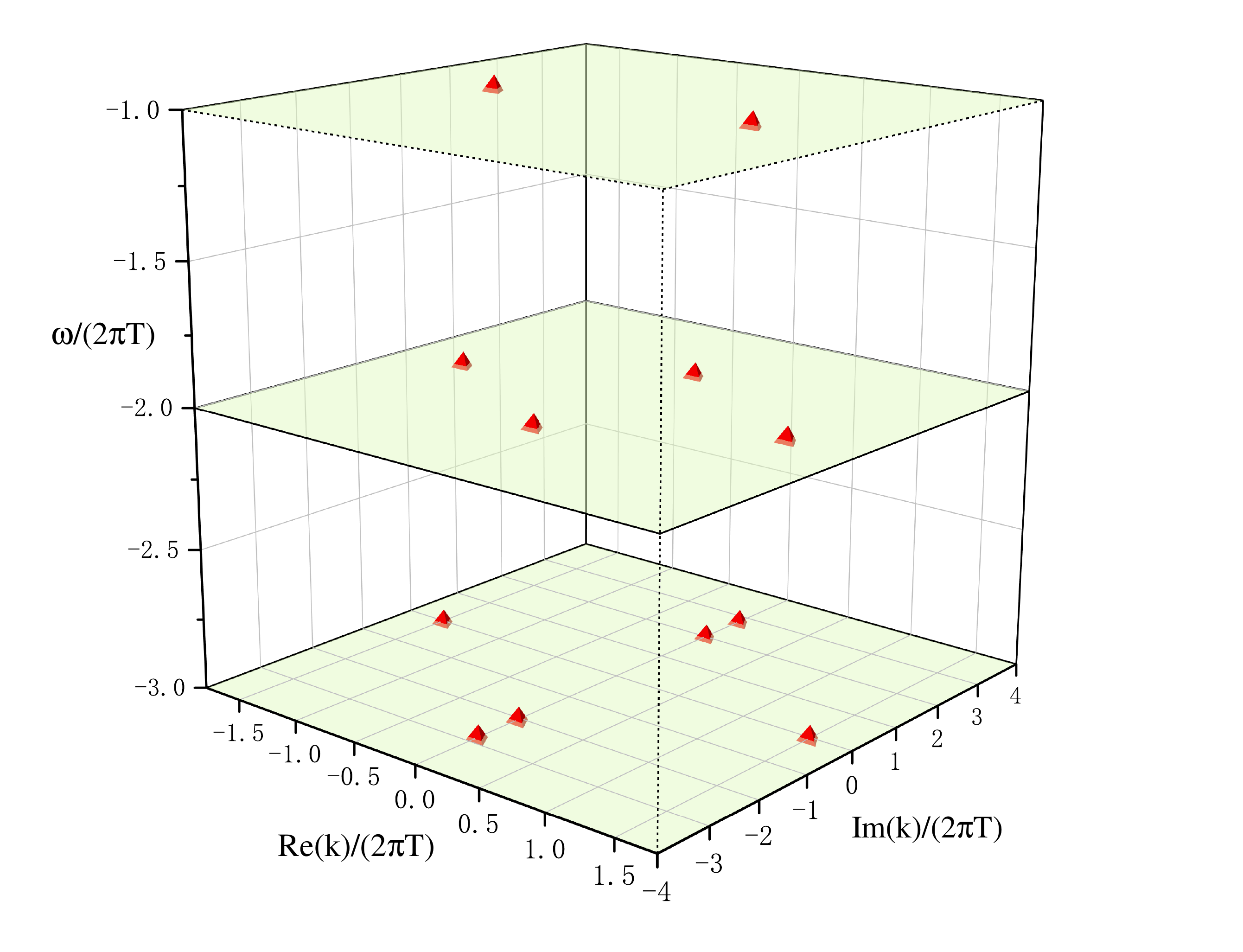}}
\centerline{(b)}
\end{minipage}
\vfill
\caption{\label{fig:Figure4} The first three order pole-skipping points of Maxwell vector mode $A_y$ in the anisotropic system near Lifshitz points . (a) $\theta=9$, $\zeta=-1$, $z=\phi=1$; (b) $\theta=4$, $\zeta=-1$, $z=\phi=-1$.  }
\end{figure}
\bea
&&\mathfrak{w}_{\star}=-i,\nonumber\\
&&\mathfrak{k}^{2}_{x\star}=\frac{2(\phi-2\kappa\zeta-1)r_+^{2z-2\phi-1}}{\vert\delta_0\vert}.
\eea
When $z=\phi=1$ and $\theta=0$, the parameter $\kappa=0$ from eq.\eqref{eq:29}. The metric \eqref{eq:3} recovers the black brane solution in $AdS_4$. The pole-skipping point become
\bea
\mathfrak{w}_{\star}=-i,\quad \mathfrak{k}^{2}_{x\star}=0.
\eea
This result is the same as (1.7b) in ref.\cite{Makoto2} (horizon radius $r_+=1$). We also evaluate the higher special points
\bea
\omega_n=-i2\pi Tn,\qquad k^2=k^2_n,\qquad {\rm det}\mathcal{M}^{(n)}(\omega,k^2)=0.
\eea
The first few elements of this matrix have showed in appendix A.1.3. Then we calculate the first three order pole-skipping points and plot them in the figure~\ref{fig:Figure4}. \\
\textbf{(a)} $z=\phi=1$, $\zeta=-1$, $\theta=9$
\bea
\mathfrak{w}_{\star1}&=&-i,\; \quad \mathfrak{k}_{x\star1}=\pm1.627;\nonumber\\
\mathfrak{w}_{\star2}&=&-2i,\quad \mathfrak{k}_{x\star2,1}=\pm2.238i,\ \mathfrak{k}_{x\star2,2}=\pm2.374;\nonumber\\
\mathfrak{w}_{\star3}&=&-3i,\quad \mathfrak{k}_{x\star3,1}=\pm2.879i,\ \mathfrak{k}_{x\star3,2}=\pm2.969,\ \mathfrak{k}_{x\star3,3}=\pm3.356i.
\eea
\textbf{(b)} $z=\phi=-1$, $\zeta=-1$, $\theta=4$
\bea
\mathfrak{w}_{\star1}&=&-i,\; \quad \mathfrak{k}_{x\star1}=\pm0.994;\nonumber\\
\mathfrak{w}_{\star2}&=&-2i,\quad \mathfrak{k}_{x\star2,1}=\pm1.285,\ \mathfrak{k}_{x\star2,2}=\pm1.695i;\nonumber\\
\mathfrak{w}_{\star3}&=&-3i,\quad \mathfrak{k}_{x\star3,1}=\pm1.494,\ \mathfrak{k}_{x\star3,2}=\pm2.028i,\ \mathfrak{k}_{x\star3,3}=\pm2.820i.
\eea

\begin{figure}[htp]
	\begin{centering}
		\includegraphics[scale=1]{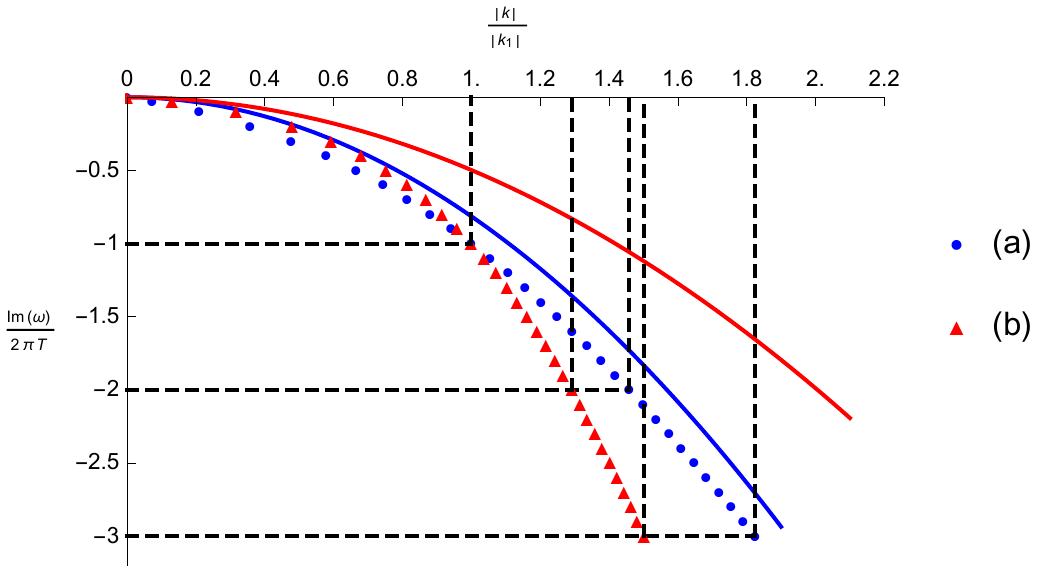}
		\par\end{centering}
	\caption{\label{fig:Figure5} Solid lines show the diffusive hydrodynamic dispersion relation\eqref{eq:17}. The intersections of the dashed lines correspond to the first three order pole-skipping points of Maxwell vector mode $A_y$ in the anisotropic system near Lifshitz points. Blue dots and red triangles show the curve fitting of pole-skipping points represent the data which we choose (a) $\theta=9$, $\zeta=-1$, $z=\phi=1$; (b) $\theta=4$, $\zeta=-1$, $z=\phi=-1$.}
\end{figure}

\indent The hydrodynamic pole corresponding to the diffusion of charge with the small-$k$ dispersion relation is given by\cite{Iqbal}
\bea
\label{eq:17}
\omega(k)=-iD_ck^2+\dots,
\eea
where the charge diffusion constant is given by $D_{c,\alpha}=-\frac{L}{\Delta_\chi}\frac{r^{\theta-z}}{g_{\alpha\alpha}(r)}\vert_{r_{+}}$, $\Delta_\chi$ is the scaling dimension of the charge susceptibility \cite{Inkof}. The momenta $k_n$ are complex numbers, so we perform the complex hydrodynamic analyse. We plot first three order pole-skipping points of $(\mathfrak{w}_{n}, \vert\mathfrak{k}_{n}\vert)$  and the hydrodynamic dispersion relation in terms of dimensionless variables $\frac{\omega}{2\pi T}$ and $\frac{\vert k\vert}{2\pi T}$  in figure~\ref{fig:Figure5}. They fit well in small $k$ limit. So this is a general conclusion in which hydrodynamic pole corresponds to the diffusion of both momentum and charge.
\subsubsection{Maxwell scalar mode}
\qquad The gauge-invariant variables of the Maxwell scalar mode
\bea
&&\mathfrak{A}_{t}=A_t+\frac{\omega}{k_x}A_{x},\nonumber\\
&&\mathfrak{A}_{r}=A_r-\frac{1}{ik_x}A'_{x}.
\eea
We consider the linear perturbation $A_{x}=a_xe^{-i\omega v+i k_x x}$. Combining the Maxwell scalar equations by gauge-invariant variables, we obtain
\bea
&&\mathfrak{A}'_{r}+\bigg(\frac{f'(r)}{f(r)}-\frac{2i\omega L r^{z-1}}{f(r)}-\frac{k^2 L r^{2\phi-z-1}}{i\omega}-(z-2\kappa\zeta-\phi)r^{-1}\bigg)\mathfrak{A}_{r}\nonumber\\
&& +\frac{L}{f(r)}\bigg((2\kappa\zeta+\phi-1) r^{z-2}-\frac{k^2_xr^{2\phi-z-2}}{i\omega}\bigg)\mathfrak{A}_{t}=0.
\eea
We use the approximation $f(r)\sim f'(r)(r-r_+)$ near the horizon radius $r=r_+$, then the field equation near $r=r_+$ is given as
\bea
\mathfrak{A}'_{r}+\frac{1-i\mathfrak{w}}{r-r_{+}}\mathfrak{A}_{r}+\frac{L}{f'(r)}\bigg((2\kappa\zeta+\phi-1) r^{z-2}-\frac{2\pi T\mathfrak{k}^{2}_xr^{2\phi-z-2}}{i\mathfrak{w}}\bigg)\bigg|_{r=r_+}\frac{\mathfrak{A}_{t}}{r-r_{+}}=0.
\eea
Making the coefficients $(1-i\mathfrak{w})$ and $\frac{L}{f'(r)}\bigg((2\kappa\zeta+\phi-1) r^{z-2}-\frac{2\pi T\mathfrak{k}^{2}_xr^{2\phi-z-2}}{i\mathfrak{w}}\bigg)\bigg|_{r=r_+}$ vanish, we obtain the first order special point about Maxwell scalar mode
\bea
&&\mathfrak{w}_{\star}=-i,\nonumber\\
&&\mathfrak{k}^{2}_{x\star}=\frac{2(\phi+2\kappa\zeta-1)r_+^{3z-2\phi-1}}{\vert\delta_0\vert}.
\eea
When $z=\phi=1$ and $\theta=0$, the result recovers the Maxwell scalar pole-skipping point corresponding to (1.7c) in ref.\cite{Makoto2} (horizon radius $r_+=1$).
\bea
\mathfrak{w}_{\star}=-i,\quad \mathfrak{k}^{2}_{x\star}=0.
\eea

\section{Lifshitz black hole with linear axion fields and hyperscaling violating factor}
\label{sec:Lifshitz}
\qquad In this section, we consider the background of Lifshitz black holes with linear axion fields and hyperscaling violating factor, and we will compute the lower half-plane pole-skipping points in the axion field. The general action is given by \cite{Ge2,Ge1,Ge3,Chen}
\bea
S=\frac{1}{16\pi G_{d+2}}\int d^{d+2}x\sqrt{-g}\big[R+V(\phi)-\frac{1}{2}(\partial\phi)^2-\frac{1}{4}\sum^{n}_{{\rm i}=1}Z_{\rm i}(\phi)F^2_{({\rm i})}-\frac{1}{2}\sum^{d}_{{\rm j}}Y(\phi)(\partial\chi_{\rm j})^2\big],\nonumber\\
\eea
where $Z_{\rm i}(\phi)=e^{\lambda_{\rm i} \phi}$ and $Y(\phi)=e^{-\lambda_2 \phi}$, $R$ is the Ricci scalar, and $\chi_{\rm i}$ is $d$-massless linear axions. $F^{(1)}_{rt}$ is an auxiliary gauge field, and  $F^{(2)}_{rt}$ is a gauge field analogous to that of a Maxwell field. The black hole solution is given by \cite{Ge2,Ge1,Chen}
\bea
\label{eq:18}
ds^2&=&r^{-\frac{2\theta}{d}}\big(-r^{2z}f(r)dt^2+\frac{dr^2}{r^2f(r)}+r^2d \vec{x}^2_d\big),
\eea
with the notations
\bea
f(r)&=&1-\frac{m}{r^{d+z-\theta}}+\frac{Q^2}{r^{2(d+z-\theta-1)}}-\frac{\beta^2}{r^{2z-2\theta/d}},\nonumber\\
F_{(1)rt}&=&Q_1\sqrt{2(z-1)(z-d-\theta)}r^{d+z-\theta-1},\nonumber\\
F_{(2)rt}&=&Q_2\sqrt{2(d-\theta)(z-\theta+d-2)}r^{-(d+z-\theta-1)},\nonumber\\
\lambda_1&=&-\frac{2d-2\theta+\frac{2\theta}{d}}{\sqrt{2(d-\theta)(z-1-\theta/d)}},\nonumber\\
\lambda_2&=&\sqrt{2\frac{z-1-\theta/d}{d-\theta}},\nonumber\\
e^{\phi}&=&r^{\sqrt{2(d-\theta)(z-1-\theta/d)}} ,\nonumber\\
V(\phi)&=&(z+d-\theta-1)(z+d-\theta)r^{2\theta/d},\nonumber\\
\chi_{\rm i}&=&\beta x^a,\qquad {\rm i}\in\{1,d\},\quad a\in\{x,y\dots\}.
\eea
The mass $m$ in terms of $r_H$ is
\bea
m=r_H^{d+z-\theta}+Q^2_2r_H^{2-d-z-\theta}-\beta^2r_H^{d-z-\theta+2\theta/d}.
\eea
The Hawking temperature is given by
\bea
T=\frac{r^{z+1}_H f'(r_H)}{4\pi}=\frac{m(2+z-\theta)r_H^{\theta-2}+\beta^2(2z-\theta)r_H^{\theta-2}+2Q^2(\theta-z-1)r_H^{2\theta-z-2}}{4\pi}.
\eea
We consider $d=2$ dimensional case in the following. For simplicity of caculation, we use the Eddington-Finkelstein (EF) coordinates. By putting the tortoise coordinate $dr_{\ast}=\frac{1}{r^{z+1}f(r)}dr$ and $v=t-r_{\ast}$ into the metric \eqref{eq:18}, we obtain
\bea
ds^2&=&r^{2z-\theta}f(r)dv^2+2r^{z-\theta-1}dvdr+r^{2-\theta}d \vec{x}^2_2.
\eea
We just want to find lower half-plane pole-skipping points which also locate at Matsubara frequencies  $\mathfrak{w}_{n}:=\frac{\omega_n}{2\pi T}=-in$ in this background. So we only consider the linear perturbation of axion field $\chi(r)=\beta x+\bar{\chi}(r)e^{-i\omega t+i k_x x}$ in this section. We choose the parameter $\beta=0.1$ which is small enough in the following. We take the equation of motion of axion field as a probe. Then we write down the equation of motion for axions \cite{Ge1,Chen,Gouteraux}
\bea
\nabla_{\mu}\big(Y(\phi)\nabla^{\mu}\chi\big)=0.
\eea
We obtain
\bea
\label{eq:20}
&&\chi''(r)+\bigg(\frac{f'(r)}{f(r)}-\frac{2i\omega r^{-z-1}}{f(r)}+(\theta-z-3+\delta)r^{-1}\bigg)\chi'(r)\nonumber\\
&& +\frac{1}{f(r)}\bigg(i\omega(\theta-2+\delta) r^{-z-2}-k^2_xr^{-4}\bigg)\chi(r)=0,
\eea
where a prime symbol denotes derivative with respect to $r$ with the notation $\delta=\vert 2-2z+\theta\vert$.  We use the approximation $f(r)\sim f'(r_H)(r-r_H)$ near the horizon $r=r_H$. The field equation near $r=r_H$ of $\chi(r)$ is given as
\bea
\label{eq:19}
&&\chi''+\frac{1}{f'(r_H)}\bigg(f'(r_H)-2i\omega r_H^{-z-1}+f'(r_H)(r-r_H)(\theta-z-3+\delta)r_H^{-1}\bigg)\frac{\chi'}{(r-r_H)}\nonumber\\
&& +\frac{1}{f'(r_H)}\bigg(i\omega(\theta-2+\delta) r_H^{-z-2}-k^2_x r_H^{-4}\bigg)\frac{\chi}{(r-r_H)}=0.
\eea
We define $\mathfrak{w}=\frac{\omega}{2\pi T}$, and $\mathfrak{k}_{x}=\frac{k_{x}}{2\pi T}$, then equation \eqref{eq:19} becomes
\bea
&& \chi''+\frac{1-i\mathfrak{w}}{r-r_H}\chi'-\bigg(\frac{i\mathfrak{w}}{2}r_H^{-1}-4\mathfrak{k}_x^{2}\pi^2T^2r_H^{-4}\bigg)\frac{\chi}{r-r_H}=0.
\eea
For a generic $(\mathfrak{w},\mathfrak{k}_x)$, the equation has a regular singularity at $r=r_H$. So we can also solve it by a power series expand around $r=r_H$
\bea
\chi(r)=(r-r_H)^\lambda \sum^\infty_{p=0}\chi_p(r-r_H)^p.
\eea
At the lowest order, we can obtain the same indicial equation $\lambda(\lambda-i\mathfrak{w})=0$ as the result of the section~\ref{sec:Aniso}. The two solutions are also
\bea
\lambda_1=0,\qquad \lambda_2=i\mathfrak{w}.
\eea
We choose $i\mathfrak{w}=1$ and appropriate $\mathfrak{k}_x$ making the singularity in front of $\chi'$ and $\chi$ terms vanish, i.e., we make  $(1-i\mathfrak{w})$ and $\bigg(\frac{i\mathfrak{w}}{2}r_H^{-1}-4\mathfrak{k}_x^{2}\pi^2T^2r_H^{-4}\bigg)$ become 0. Then we find the location of the special point about $\chi(x)$ field equation
\bea
&&\mathfrak{w}_{\star}=-i,\nonumber\\
&&\mathfrak{k}^{2}_{x\star}=\frac{2(\theta-2+\delta)}{-2Q^2(1+z-\theta)r_H^{2\theta-4}-\beta^2(\theta-2z)r_H^{\theta-2}-m(\theta-z-2)r_H^{\theta+z-4}}.
\eea
When $\theta=0$, $z=1$, $Q=0$ and $\beta=0$, the metric \eqref{eq:18} recovers the black brane solution in $AdS_4$. The pole-skipping point become
\bea
\mathfrak{w}_{\star}=-i,\quad \mathfrak{k}^{2}_{x\star}=-\frac{4}{3}.
\eea
This result is the same as (1.7a) in ref.\cite{Makoto2} (horizon radius $r_H=1$). We extend pole-skipping phenomenon at higher Matsubara frequencies $\omega_n=-i2\pi Tn$, and the method has been showed in section~\ref{sec:Aniso}. We expand the $\chi(r)$ with a Taylor series
\bea
\label{eq:21}
\chi(r)=\sum^\infty_{p=0}\chi_p(r-r_H)^p=\chi_0+\chi_1(r-r_H)+\dots.
\eea
We insert \eqref{eq:21} into \eqref{eq:20}, then a series of perturbed equation in the order of $(r-r_H)$ can be denoted as
\bea
\label{eq:22}
S=\sum^\infty_{p=0}S_p(r-r_H)^p=S_0+S_1(r-r_H)^p+\dots.
\eea
We write down the first few equation $S_p=0$ in the expansion of \eqref{eq:22}
\bea
&&0=M_{11}(\omega,k^2)\chi_0+(2\pi T-i\omega)\chi_1,\nonumber\\
&&0=M_{21}(\omega,k^2)\chi_0+M_{22}(\omega,k^2)\chi_1+(4\pi T-i\omega)\chi_2,\nonumber\\
&&0=M_{31}(\omega,k^2)\chi_0+M_{32}(\omega,k^2)\chi_1+M_{33}(\omega,k^2)\chi_2+(6\pi T-i\omega)\chi_3,
\eea
\begin{figure}[!t]
\begin{minipage}{0.48\linewidth}
\centerline{\includegraphics[width=7.5cm]{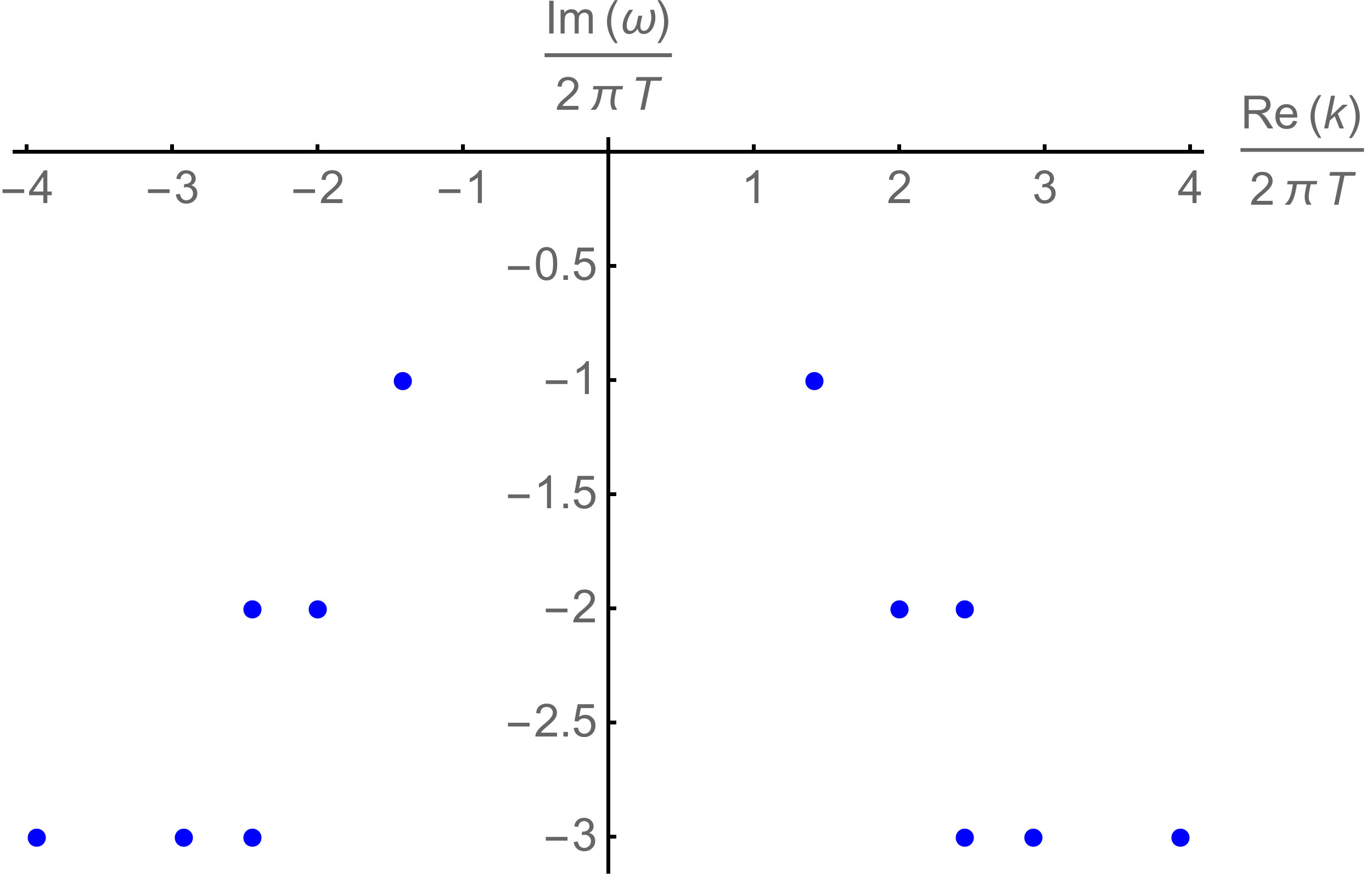}}
\centerline{(a)}
\end{minipage}
\hfill
\begin{minipage}{0.48\linewidth}
\centerline{\includegraphics[width=7.5cm]{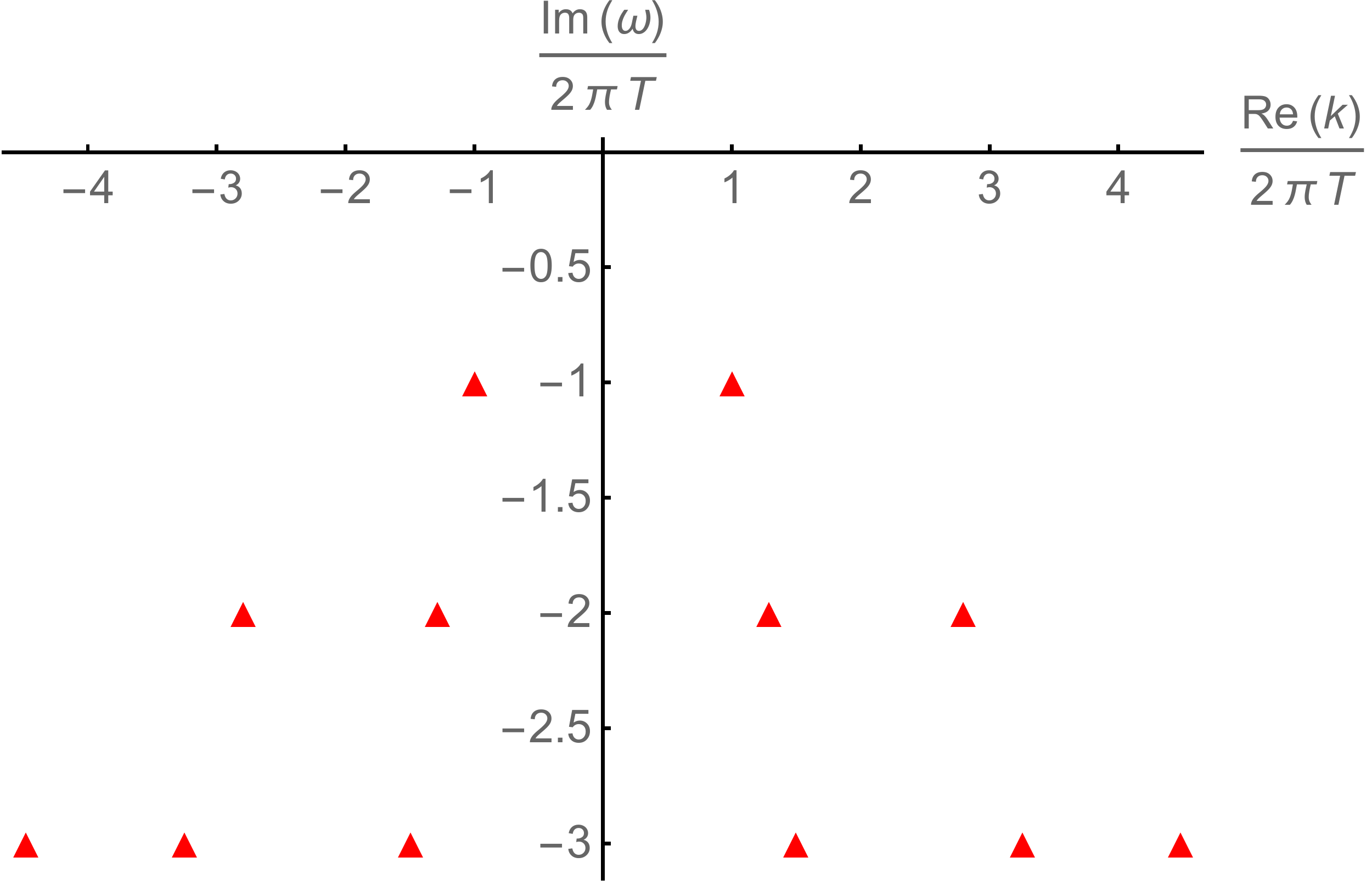}}
\centerline{(b)}
\end{minipage}
\vfill
\caption{\label{fig:Figure6} The first three order pole-skipping points of axion field $\chi(r)$ in Lifshitz spacetime with linear axion fields and hyperscaling violating factor. (a) $z=2$, $\theta=3$; (b) $z=2.5$, $\theta=2.5$.}

\end{figure}
In order to find an ingoing solution, we should solve a set of linear equations of the form
\bea
\label{eq:27}
\mathcal{M}^{(n)}(\omega,k^2)\cdot \chi\equiv\left(\begin{array}{ccccc}
    M_{11} & (2\pi T-i\omega) & 0    & 0  &\dots\\
    M_{21} & M_{22}& (4\pi T-i\omega)& 0   &\dots\\
    M_{31} & M_{32}&  M_{33} &(6\pi T-i\omega) &\dots\\
    \dots   &  \dots&  \dots  &\dots   &\dots\\
\end{array}\right)\left(\begin{array}{ccccc}
    \chi_0\\
   \chi_1\\
    \chi_2 \\
    \dots \\
\end{array}\right)=0
\eea
The locations of special points $(\omega_n, k_n)$  can be easily extracted from determinant of the $(n\times n)$ matrix $\mathcal{M}^{(n)}(\omega,k^2)$ constructed by the first $n$ equations:
\bea
\omega_n=-i2\pi Tn,\qquad k^2=k^2_n,\qquad {\rm det}\mathcal{M}^{(n)}(\omega,k^2)=0.
\eea
The first few elements of this matrix have showed in Appendix A.2. The first three order pole-skipping points are shown in figure~\ref{fig:Figure6}. We choose the charges $Q_1=Q_2=1$ in the following. \\
\textbf{(a)} $z=2$, $\theta=3$
\bea
\mathfrak{w}_{\star1}&=&-i,\; \quad \mathfrak{k}_{\star1}=\pm1.414;\nonumber\\
\mathfrak{w}_{\star1}&=&-2i,\quad \mathfrak{k}_{\star2,1}=\pm2.000,\ \mathfrak{k}_{\star2,2}=\pm2.449;\nonumber\\
\mathfrak{w}_{\star1}&=&-3i,\quad \mathfrak{k}_{\star3,1}=\pm2.449,\ \mathfrak{k}_{\star3,2}=\pm2.922,\ \mathfrak{k}_{\star3,3}=\pm3.932.
\eea
\textbf{(b)} $z=2.5$, $\theta=3$
\bea
\mathfrak{w}_{\star1}&=&-i,\; \quad \mathfrak{k}_{\star1}=\pm0.999;\nonumber\\
\mathfrak{w}_{\star2}&=&-2i,\quad \mathfrak{k}_{\star2,1}=\pm1.286,\ \mathfrak{k}_{\star2,2}=\pm2.795;\nonumber\\
\mathfrak{w}_{\star3}&=&-3i,\quad \mathfrak{k}_{\star3,1}=\pm1.495,\ \mathfrak{k}_{\star3,2}=\pm3.252,\ \mathfrak{k}_{\star3,3}=\pm4.480.
\eea
 We depict the first three order pole-skipping points and hydrodynamic dispersion relation \eqref{eq:10} in figure~\ref{fig:Figure7}. They fit well at small $\omega$ and $k$.
 \begin{figure}[htp]
	\begin{centering}
		\includegraphics[scale=1]{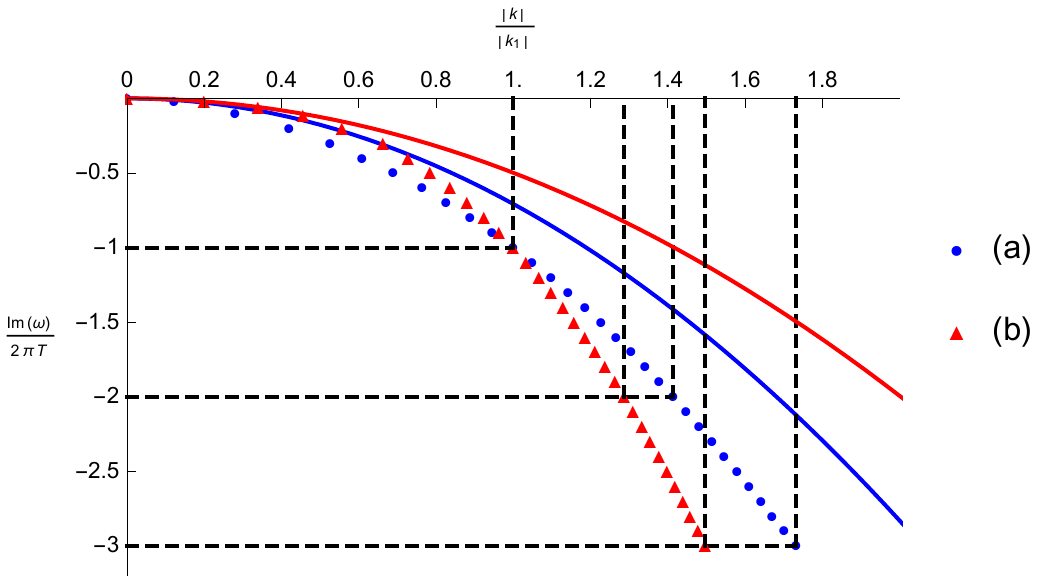}
		\par\end{centering}
	\caption{\label{fig:Figure7} Solid lines show the diffusive hydrodynamic dispersion relation\eqref{eq:10}. The intersections of the dashed lines correspond to the first three order pole-skipping points of axion field $\chi(r)$.  Blue dots and red triangles show the curve fitting of pole-skipping points represent the data which we choose  (a) $z=2$, $\theta=3$; (b) $z=2.5$, $\theta=3$.)}
\end{figure}
\section{AdS$_2\times\mathbb{R}^{d-1}$ geometry}
\label{sec:AdS}
\qquad We are interested in the locations of pole-skipping points AdS$_2\times\mathbb{R}^{d-1}$  and we want to see if different boundary conditions have an impact on the location of pole-skipping points. The AdS$_2\times\mathbb{R}^{d-1}$ geometry is given as\cite{Thomas}.
\bea
\label{eq:23}
ds^2=\frac{R_2^2}{\zeta^2}\bigg(-(1-\frac{\zeta^2}{\zeta_0^2})d\tau^2+\frac{d\zeta^2}{1-\frac{\zeta^2}{\zeta^0}}\bigg)+\frac{r_*^2}{R^2}dx^2+\frac{r_*^2}{R^2}dy^2,
\eea
where a temperature $T=\frac{1}{2\pi\zeta_0}$. $r_*$ is a length scale of charge $Q$ which has dimension of $[L]^{d-1}$, $Q\equiv\sqrt{\frac{d}{d-2}}r_*^{d-1}$. Finite value parameter $\zeta_0$ is the horizon radius. $R_2$ is the curvature radius of AdS$_2$ and $R$ is the curvature radius of AdS. Using the Eddington-Finkelstein (EF) coordinates, the metric \eqref{eq:23} become
\bea
ds^2=-\frac{R_2^2}{\zeta^2}\bigg((1-\frac{\zeta^2}{\zeta_0^2})dv^2+2\frac{R_2^2}{\zeta^2}dv d\zeta\bigg)+\frac{r_*^2}{R^2}d\vec{x}^2.
\eea
We consider the scalar field
\bea
-\frac{1}{\sqrt{-g}}\partial_\mu(\sqrt{-g}g^{\mu\nu}\partial_\nu (\phi(\zeta)e^{-i\omega v+ikx}))=0.
\eea
We obtain
\bea
\label{eq:28}
\phi''+\frac{2(\zeta-i\omega \zeta_0^2)}{\zeta^2-\zeta_0^2}\phi'+\frac{4k^2\zeta_0^2}{R_2^2r_*\zeta^2(\zeta^2-\zeta_0^2)}\phi=0.
\eea
Using the method we have mentioned in section~\ref{sec:Aniso}, we obtain the pole-skipping points located at $\omega=-n2i\pi T (n\; \text{is integer})$. The first three order pole-skipping points are
\bea
&&\mathfrak{w}_{\star1}=-i,\; \quad \mathfrak{k}^{2}_{\star1}=0;\nonumber\\
&&\mathfrak{w}_{\star2}=-2i,\quad \mathfrak{k}^{2}_{\star2,1}=0,\quad \mathfrak{k}^{2}_{\star2,2}=\frac{\zeta_0^2R_2^2r_*}{2};\nonumber\\
&&\mathfrak{w}_{\star3}=-3i,\quad \mathfrak{k}^{2}_{\star3,1}=0,\quad \mathfrak{k}^{2}_{\star3,2}=\frac{\zeta_0^2R_2^2r_*}{2},\quad \mathfrak{k}^{2}_{\star3,3}=\frac{3\zeta_0^2R_2^2r_*}{2};\nonumber\\
&&\mathfrak{w}_{\star4}=-4i,\quad \mathfrak{k}^{2}_{\star4,1}=0,\quad \mathfrak{k}^{2}_{\star4,2}=\frac{\zeta_0^2R_2^2r_*}{2},\quad \mathfrak{k}^{2}_{\star4,3}=\frac{3\zeta_0^2R_2^2r_*}{2},\quad \mathfrak{k}^{2}_{\star4,4}=3\zeta_0^2R_2^2r_*.
\eea
\begin{figure}[!t]
\begin{minipage}{0.48\linewidth}
\centerline{\includegraphics[width=7.5cm]{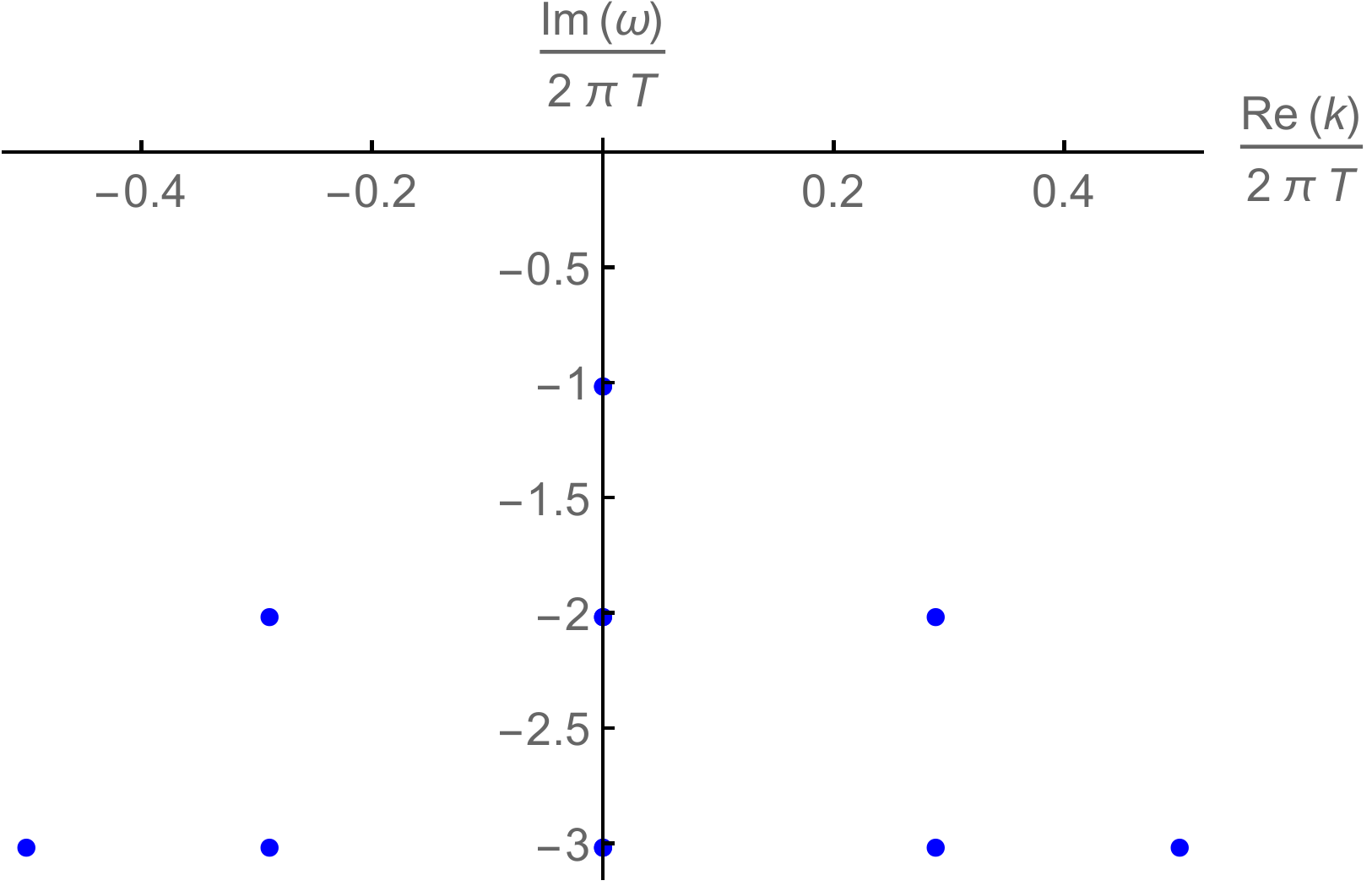}}
\centerline{(a)}
\end{minipage}
\hfill
\begin{minipage}{0.48\linewidth}
\centerline{\includegraphics[width=7.5cm]{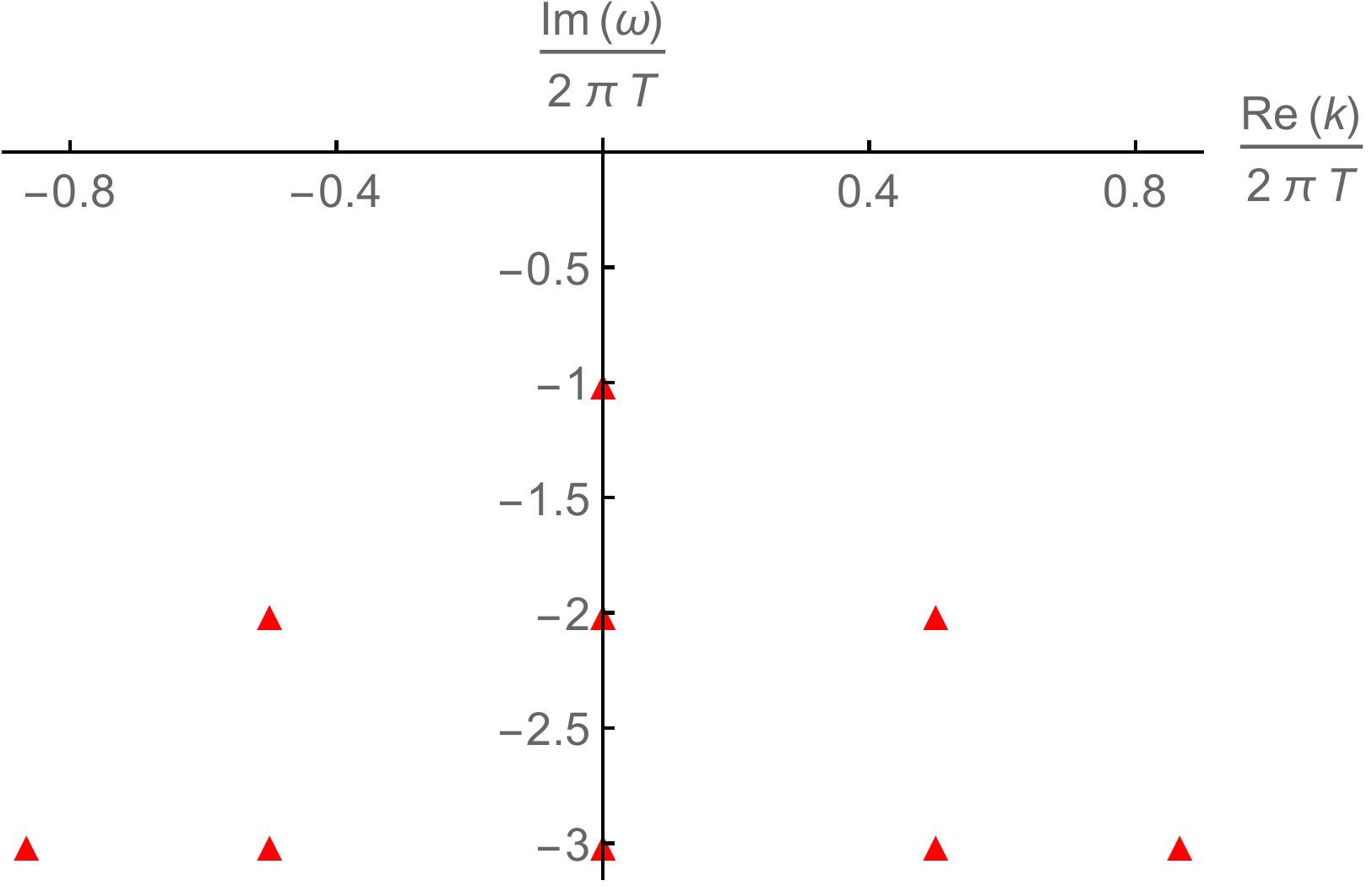}}
\centerline{(b)}
\end{minipage}
\vfill
\caption{\label{fig:Figure8} The first three order pole-skipping points of scalar field $\phi(\zeta)$ in AdS$_2\times\mathbb{R}^{d-1}$ geometry (a) $d=3$, $\zeta_0=1$, $R_2=-\frac{1}{\sqrt{6}}$; (b) $d=2$, $\zeta_0=1$, $R_2=-\frac{1}{\sqrt{2}}$.}

\end{figure}
The first few elements of this matrix have showed in appendix A.3. We plot these points in figure~\ref{fig:Figure8}. The result of first order pole-skipping point is the same as (1.7a) in ref.\cite{Makoto2}.\\
In the standard quantization, we set the conformal dimension $\triangle=\frac{1}{2}-\nu$ and the retarded Green's function is given as
\bea
\mathcal{G}_R=(4\pi T)^{2\nu}\frac{\Gamma(-2\nu)\Gamma(\frac{1}{2}+\nu-\frac{i\omega}{2\pi T}+iqe_d)\Gamma(\frac{1}{2}+\nu-iqe_d)}{\Gamma(2\nu)\Gamma(\frac{1}{2}-\nu-\frac{i\omega}{2\pi T}+iqe_d)\Gamma(\frac{1}{2}-\nu-iqe_d)}
\eea
in ref.\cite{Thomas}.  For convenience to compare with the Green's function, we consider the charge of background $Q$ and charge of external field $q$ equal to 0. So the length scale of charge $r_*=0$ and we do not need to consider the degrees of freedom of  momenta ($k_n=0$). We focus on the locations of frequencies of pole-skipping points.\\
\indent The zeros of the Green's function are: $\frac{1}{2}+\nu-\frac{i\omega}{2\pi T}=-{\rm i}, \quad {\rm i}=0,1,2...$,\nonumber\\
\indent The poles of the Green's function are: $\frac{1}{2}-\nu-\frac{i\omega}{2\pi T}=-{\rm j}, \quad {\rm j}=0,1,2...$,\nonumber\\
 where $\nu=\frac{1}{2}$. According to these two equations, the pole-skipping points are located at $\omega=-n2i\pi T (n\; \text{is integer})$ which are the same as the results from near horizon analysis.\\
\indent The positions of pole-skipping points will not change whether in the case of standard quantization or alternative quantization\cite{Yongjun1}. If we consider the alternative quantization, we just take the inverse of $\mathcal{G}_R$ and replace $\triangle\rightarrow 1-\triangle$. After that, we obtain the same special points as choosing the standard quantization.

\section{Rindler geometry}
\label{sec:Rindler}
\qquad  The Rindler metric describes the near horizon geometry of finite-temperature black hole. The ubiquity of the Rindler horizon could be not enough to determine a pole-skipping in the corresponding Green's functions. However we can use the near horizon analysis to obtain ``special points'' making the equation of motion have two incoming solutions. The Rindler metric is given as\cite{Robert}
\bea
ds^2=-K^2\rho^2dt^2+d\rho^2+\frac{1}{4K^2}dx^2+\frac{1}{4K^2}dy^2,
\eea
where $\rho$ is  proper distance and $K$ is surface gravity. The Eddington coordinate of Rindler metric can be written as
\bea
ds^2=-K^2\rho^2dv^2+2K\rho dvd\rho+\frac{1}{4K^2}dx^2+\frac{1}{4K^2}dy^2.
\eea
\subsection{sound modes}
\qquad Considering the sound modes of metric perturbation
\bea
h_{vv}=e^{-i\omega v+ikx}h_{vv}(\rho),\quad h_{vx}=e^{-i\omega v+ikx}h_{vx}(\rho),\\
h_{xx}=e^{-i\omega v+ikx}h_{xx}(\rho),\quad h_{yy}=e^{-i\omega v+ikx}h_{yy}(\rho).
\eea
Substituting these perturbations into the linearized Einstein equation
\bea
R_{\mu\nu}-\frac{1}{2}R g_{\mu\nu}=0.
\eea
We consider the $vv$ component near the horizon
\bea
k^2h_{vv}+2k(\omega-iK)h_{vx}+\omega(\omega-iK)(h_{xx}+h_{yy}).
\eea
The lower-half $\omega$-plane of ``special point'' is located at
\bea
\omega_{\star}=i2\pi T,\quad k^{2}_{\star}=0.
\eea
We obtain the Lyapunov exponent $\lambda_L=2\pi T$ from ``special point'' and it represent the upper-half $\omega$-plane pole-skipping point corresponding to the maximal chaos. This result is the same as the one in Schwarzschild-AdS and Lifshitz geometries.
\subsection{scalar field}
\qquad For a Klein-Gordon equation
\bea
-\frac{1}{\sqrt{-g}}\partial_\mu(\sqrt{-g}g^{\mu\nu}\partial_\nu (\phi(\rho)e^{-i\omega v+ikx}))=0.
\eea
We obtain
\bea
\label{eq:24}
\phi''+\frac{K-2i\omega}{K}\frac{\phi'}{\rho}-4k^2K^2\phi=0.
\eea
We calculate the higher order of the ``special points'' about scalar field by using the method in the section~\ref{sec:Aniso}
\bea
&&\omega_{n\ast}=-in\pi T,\quad k^2_{n\ast}=C_n.\quad n=1,2,3\dots,\quad  \text{$C_n$ are arbitrary constant.}
\eea
So we just determine one parameter $\omega$ to ensure two independent incoming wave rather than two parameters $\omega$ and $k$ in Rindler horizon. The equation \eqref{eq:24} have singularity at $\rho=0$. The solution can be written as a Taylor series
\bea
\phi(\rho)=\rho^\lambda\sum^\infty_{n=0}\phi_{n}\rho^n.
\eea
We can obtain the indicial equation $\lambda(\lambda-\frac{i\omega}{\pi T})=0$ and two solutions are
\bea
\lambda_1=0, \quad \lambda_2=\frac{i\omega}{\pi T}.
\eea
If we choose $i\omega=\pi T$ to make the coefficient $(\frac{K-2i\omega}{K})$ become 0, the two solutions also become
\bea
\lambda_1=0, \quad \lambda_2=1.
\eea
In the absence of the boundary condition, it maybe not enough to ensure a ``pole-skipping'' in the corresponding Green's function in the local Rindler structure. But we can obtain ``special points'' with two incoming waves from near horizon in this section. Two incoming solutions correspond to nonunique of the Green's function in physical meaning from the perspective of holography. We think this phenomenon still reflects the uncertainty of physics and it is a meaningful topic in the future.
\section{Discussion and Conclusion}
\qquad In summary, we show that near horizon analysis is a general and simpler method to calculate pole-skipping points. It can not only be applied to AdS spacetime, but also to Lifshitz and Rindler geometry. The pole-skipping points are related to the hydrodynamic dispersion relations in Lifshitz geometry. we calculate the lower-half $\omega$-plane pole-skipping points of tensor and Maxwell fields in the anisotropic system near Lifshitz points. The frequencies  of these special points are located at $\mathfrak{w}_{n}=-in$ with $n$ integers. To verify our results, we compare them with outcomes of the AdS$_4$ black brane solution in ref.\cite{Makoto1,Makoto2} when setting $z=1$, $\theta=0$ and $\phi=1$. They are the same at the first order pole-skipping point ($\mathfrak{w}_{n}=-i$). The momenta $k_n$ of higher order pole-skipping points are in the complex $k$-plane and we plot the dispersion relations in terms of dimensionless variables $\frac{\omega}{2\pi T}$ and $\frac{\vert k\vert}{2\pi T}$ . We find that the hydrodynamic dispersion relations pass through pole-skipping points $(\mathfrak{w}_{n}, \vert\mathfrak{k}_{n}\vert)$ at very small momentum and frequency. For the hydrodynamic analysis, we just consider the linearized perturbation along the $x$-direction of tensor mode which translational symmetry is not broken and momentum is not dissipated $(a=0)$ and of Maxwell vector mode which current is conserved.\\
\indent In the background of the Lifshitz black hole with linear axion field and hyperscaling violating factor, the axion field was shown with the same phenomenon. The lower half-plane pole-skipping points at negative integer (imaginary) Matsubara frequencies. When $\theta=0$, $z=1$, $Q=0$ and $\beta=0$, the metric \eqref{eq:18} recovers the AdS$_4$ black brane solution. The location of the first order pole-skipping point is the same as that of ref.\cite{Makoto1,Makoto2}. We also compare the lower half-plane pole-skipping points with hydrodynamic poles show that the hydrodynamic dispersion relation passes through pole-skipping points when $\mathfrak{w}, \mathfrak{k}\rightarrow 0$.\\
\indent We calculate the pole-skipping points of scalar field in the background of AdS$_2\times\mathbb{R}^{d-1}$. When we choose the different quantization, the locations of pole-skipping points are not change, although this phenomenon have been already found in ref.\cite{Yongjun1}. This is a general conclusion in all background.\\
\indent We cannot obtain poles-kipping points by solving Green's function directly because of lack of boundary conditions in Rindler geometry. But near the horizon, we can obtain the ``special points'' by  solving the bulk equations of motion. There are two incoming waves at these ``special points'' consist with poles-kipping points obtained by solving Green's function. \\
\indent We wish to verify that it is a general conclusion that the pole-skipping points are related to the hydrodynamic dispersion relation, especially in the anisotropic geometry along the direction which momentum is dissipated. We are also interested in why the ``pole-skipping'' points can be obtained without boundary conditions in Rindler geometry or other systems. It may have important research significance in holography.

\acknowledgments

We would like to thank Keun-Young Kim, Yan Liu, Yu-Qi Lei, and Qing-Bing Wang for helpful discussions. This work is partly supported by NSFC (No.11875184).\\

\appendix
\section{Details of near-horizon expansions}
In this appendix, we show the details of the near-horizon expansions of the equations of motion.
\subsection{Anisotropic system near Lifshitz points}
\qquad We can calculate a Taylor series solution to the tensor mode $h^x_y$, Maxwell vector mode $A_y$ equations of motion when the matrix equation \eqref{eq:25} is satisfied. For the convenience of numerical calculation, we set the horizon radius $r_+=1$ and $L=1$.  The first few elements of this matrix are shown below.
\subsubsection{Tensor mode ($a$=0)}
\qquad For the perturbation along $x$-direction of tensor mode $h^x_y$ equation \eqref{eq:4} when $a=0$. The first few elements of this matrix
\bea
M_{11}&=&-\frac{1}{2r_+}\{k^2Lr_++i\omega(\theta-2)\},\nonumber\\
M_{21}&=&-\frac{1}{2r_+^2}\{2k^2Lr_++i\omega (\theta-2)\},\nonumber\\
M_{22}&=&-\frac{1}{2r_+}\{k^2Lr_+-4\pi T\theta+i\omega(\theta+2)\}+\frac{f''(r_+)}{2L},\nonumber\\
M_{31}&=&-\frac{k^2L}{r_+^2},\nonumber\\
M_{32}&=&\frac{1}{2L r_+^{2}}\{-2L(2k^2Lr_+-4\pi T(\theta-1)+i\omega\theta)+r_+(\theta+2)f''(r_+)+r_+^2f^{(3)}(r_+)\},\nonumber\\
M_{33}&=&-\frac{1}{2r_+}\{k^2Lr_+-8\pi T(\theta-2)+i\omega(6+\theta)\}+\frac{3f''(r_+)}{2L}.\nonumber
\eea
\subsubsection{Tensor mode ($a$ $\neq$ 0)}
\qquad For the perturbation along $x$-direction of tensor mode $h^x_y$ equation \eqref{eq:26} when $a\neq0$. The first few elements of this matrix
\bea
M_{11}&=&-\frac{1}{2r_+}\{k^2Lr_+^\phi+a^2Lr_+^{2\kappa\lambda+\phi}+i\omega(1+\theta-3\phi)\},\nonumber\\
M_{21}&=&-\frac{1}{2r_+^2}\{2k^2Lr_+^\phi\phi+2a^2Lr_+^{2\kappa\lambda+\phi}(\kappa\lambda+\phi)+i\omega \phi(1+\theta-3\phi)\},\nonumber\\
M_{22}&=&-\frac{1}{2r_+L}\{L(k^2Lr_+^\phi+a^2Lr_+^{2\kappa\lambda+\phi}-4\pi T(4+\theta-4\phi)+i\omega(3+\theta-\phi))-r_+^{2-\phi}f''(r_+)\},\nonumber\\
M_{31}&=&-\frac{1}{2r_+^3}\{2a^2Lr_+^{2\kappa\lambda+\phi}(2\kappa^2\lambda^2+\phi(2\phi-1)+\kappa\lambda(4\phi-1))+\phi(2k^2Lr_+^\phi(2\phi-1)\nonumber\\
&&+i\omega(1+\theta-3\phi)(\phi-1))\},\nonumber\\
M_{32}&=&\frac{r_+^{2-\phi}}{2L}\{-2Lr_+^\phi(-4\pi T(3+\theta-4\phi)+2k^2Lr_+^\phi\phi+2a^2Lr_+^{2\kappa\lambda+\phi}(\kappa\lambda+\phi)+i\omega\phi(2+\theta-2\phi))\nonumber\\
&&+r_+^2(6+\theta-4\phi)f''(r_+)+r_+^3f^{(3)}(r_+)\},\nonumber\\
M_{33}&=&-\frac{1}{2r_+}\{k^2Lr_+^\phi+a^2Lr_+^{2\kappa\lambda+\phi}-8\pi T(6+\theta-4\phi)+i\omega(5+\theta+\phi)\}+\frac{3r_+^{1-\phi}f''(r_+)}{2L}.\nonumber
\eea
\subsubsection{Maxwell vector mode}
\qquad For the Maxwell vector $A_y$ equation of motion \eqref{eq:31}. The first few elements of this matrix
\bea
M_{11}&=&\frac{1}{2}\{-k^2-i\omega(1+2\zeta\kappa-\phi)\},\nonumber\\
M_{21}&=&\frac{1}{4}\{-2k^2\phi-i\omega z(1+2\zeta\kappa-\phi)\},\nonumber\\
M_{22}&=&\frac{1}{4}\{-k^2-4\pi T(z-2\zeta\kappa-\phi-4)-i\omega(3+2z+\zeta\kappa-\phi)+f''(1)\},\nonumber\\
M_{31}&=&\frac{1}{12}\{-2k^2\phi(2\phi-1)-z(z-1)i\omega(1+2\zeta\kappa-\phi)\},\nonumber\\
M_{32}&=&\frac{1}{12}\{-4k^2\phi-8\pi T(z+\phi-2\zeta\kappa-3)-2zi\omega(z+2\zeta\kappa+2-\phi)-(z+\phi-2\zeta\kappa-6)f''(1)\nonumber\\
&&+f^{(3)}(1)\},\nonumber\\
M_{33}&=&\frac{1}{6}\{-k^2-8\pi T(z-\zeta\kappa+\phi-6)-i\omega(4z+2\zeta\kappa-\phi+5)+3f''(1) \}.\nonumber
\eea
\subsection{Lifshitz black hole with linear axion fields and hyperscaling violating factor}
\qquad We can calculate a Taylor series solution to the equation of axion field $\chi(r)$ \eqref{eq:20} when the matrix equation \eqref{eq:27} is satisfied. For convenience , we set the horizon radius $r_H=1$. The first few elements of this matrix
\bea
M_{11}&=&\frac{1}{2}\{-k^2+i\omega (\delta+\theta-2)\},\nonumber\\
M_{21}&=&\frac{1}{4}\{4k^2-i\omega (z+2)(\delta+\theta-2)\},\nonumber\\
M_{22}&=&\frac{1}{4}\{-k^2+i\omega (\delta+\theta-2)-4\pi T(\delta+\theta-z-3)+2i\omega(z+1)+f''(1)\},\nonumber\\
M_{31}&=&\frac{1}{12}\{-20k^2+i\omega(z+2)(z+3)(\delta+\theta-2)\},\nonumber\\
M_{32}&=&\frac{1}{12}\{8k^2-2i\omega(z+2)(\delta+\theta-2)-2i\omega(z+1)(z+2)+(\delta+\theta-z-3)(8\pi T-f''(1))\nonumber\\
&&+f^{(3)}(1) \},\nonumber\\
M_{33}&=&\frac{1}{6}\{-k^2+i\omega(\delta+\theta-2)+f''(1)-8\pi T(\delta+\theta-z-3)+4i\omega(z+1)+2f''(1)\}.\nonumber
\eea
\subsection{AdS$_2\times\mathbb{R}^{d-1}$ geometry}
\qquad For the scalar field $\phi$ equation of motion \eqref{eq:28}. The first few elements of this matrix
\bea
&&M_{11}=\frac{4k^2\pi^2T^2}{R^2_2r_*},\qquad\qquad M_{21}=0,\nonumber\\
&&M_{22}=\frac{2\pi T(2k^2\pi T+R^2_2r_*(3\pi T-i\omega))}{R^2_2r_*},\nonumber\\
&&M_{31}=0,\qquad\qquad\qquad\quad M_{32}=4\pi^2T^2(6\pi T-i\omega),\nonumber\\
&&M_{33}=\frac{4\pi T(2k^2\pi T+R^2_2r_*(11\pi T-2i\omega))}{3R^2_2r_*},\nonumber\\
&&M_{41}=0,\qquad\qquad\qquad\quad M_{42}=4\pi^4T^4,\nonumber\\
&&M_{43}=2\pi^2 T^2(10\pi T-i\omega),\nonumber\\
&&M_{44}=\frac{2k^2\pi^2 T^2+24\pi^2 T^2R^2_2r_*-3i\omega\pi TR^2_2r_*}{R^2_2r_*}.\nonumber
\eea


\begin{thebibliography}{99}
\bibitem{maldacena}
J. Maldacena, \emph{The Large-N Limit of Superconformal Field Theories and Supergravity}, \emph{Int. J. Theor. Phys.} {\bf 38} (1999) 1113 [arXiv:9711200].
\bibitem{witten1}E. Witten, \emph{Anti-de Sitter space and holography}, \emph{Adv. Theor. Math. Phys.} {\bf 2} (1988) 253 [arXiv:9802150].
\bibitem{witten2}E. Witten, \emph{Anti-de Sitter space, thermal phase transition, and confinement in gauge theories}, \emph{Adv. Theor. Math. Phys.} {\bf 2} (1998)  505 [arXiv:9803131].
\bibitem{gubser}S. S. Gubser, I. R. Klebanov and A. M. Polyakov, \emph{Gauge theory correlators from noncritical string theory}, \emph{Phys. Lett. B.} {\bf428} (1998) 105 [arXiv:9802109].
\bibitem{casalderrey}J. Casalderrey-Solana, H. Liu, D. Mateos, K. Rajagopal and U. A. Wiedemann, \emph{Gauge/String Duality, Hot QCD and Heavy Ion Collisions} Cambridge Univ. Press (2014).
\bibitem{natsuume}M. Natsuume, \emph{AdS/CFT Duality User Guide}, Springer Japan, Tokyo (2015).
\bibitem{ammon}M. Ammon and J. Erdmenger, \emph{Gauge/gravity duality: Foundations and applications}, Cambridge Univ. Press (2015).
\bibitem{zaanen}J. Zaanen, Y. W. Sun, Y. Liu and K. Schalm, \emph{Holographic Duality in Condensed Matter Physics}, Cambridge Univ. Press (2015).
\bibitem{hartnoll}S. A. Hartnoll, A. Lucas and S. Sachdev, \emph{Holographic quantum matter}, The MIT Press (2018).
\bibitem{Shenker1}S. H. Shenker and D. Stanford, \emph{Black holes and the butterfly effect},  \emph{JHEP.} {\bf2014} (2014) 67 [arXiv:1306.0622].
\bibitem{Roberts}D. A. Roberts, D. Stanford and L. Susskind, \emph{Localized shocks}, \emph{JHEP.} {\bf2015} (2015) 051 [arXiv:1409.8180].
\bibitem{Shenker2}S. H. Shenker and D. Stanford, \emph{Stringy effects in scrambling}, \emph{JHEP.} {\bf2015} (2015) 132 [arXiv:1412.6087].
\bibitem{Makoto1}M. Natsuume and T. Okamura, \emph{Holographic chaos, pole-skipping, and regularity}, \emph{Progress of Theoretical and Experimental Physics}, {\bf 1} (2020) 013B07 [arXiv:1905.12014].
\bibitem{Makoto2}M. Natsuume and T. Okamura, \emph{Nonuniqueness of Green's functions at special points}, arXiv:1905.12015.
\bibitem{Grozdanov1}S. Grozdanov, K. Schalm and V. Scopelliti, \emph{Black hole scrambling from hydrodynamics}, \emph{Phys. Rev. Lett.} {\bf 120} (2018) 231601 [arXiv:1710.00921].
\bibitem{Blake}M. Blake, R. A. Davions, S. Grozdanov and H. Liu, \emph{Many-body chaos and energy dynamics in holography}, \emph{JHEP.} {\bf 2018} (2018) 035 [arXiv:1809.01169].
\bibitem{Grozdanov2}S. Grozdanov, \emph{On the connection between hydrodynamics and quantum chaos in holographic theories with stringy corrections}, \emph{JHEP.} {\bf 2019} (2019) 48 [arXiv:1811.09641].
\bibitem{BlakeDavison}M. Blake, R.A. Davison and D. Vegh, \emph{Horizon constraints on holographic Green's functions} \emph{J. High Energ. Phys.} {\bf 2020} (2020) 77 [arXiv:1904.12883].
\bibitem{Makoto3}M. Natsuume and T. Okamura, \emph{Pole-skipping with finite-coupling corrections}, \emph{Phys. Rev. D.} {\bf 100} (2019) 126012 [arXiv:1909.09168].
\bibitem{Das}S. Das, B. Ezhuthachan and A. Kundu,  \emph{Real time dynamics from low point correlators in 2d BCFT}, \emph{J. High Energ. Phys.} {\bf 2019} (2019) 141.
\bibitem{Abbasi1} N. Abbasi and J. Tabatabaei, \emph{Quantum chaos, pole-skipping and hydrodynamics in a holographic system with chiral anomaly}, \emph{J. High Energ. Phys.} {\bf 2020} (2020) 50 [arXiv:1910.13696].
\bibitem{Abbasi2} N. Abbasi and S. Tahery, \emph{Complexified quasinormal modes and the pole-skipping in a holographic system at finite chemical potential}, \emph{J. High Energ. Phys.} {\bf 2020} (2020) 76 [arXiv:2007.10024].
\bibitem{Yongjun1}Y. Ahn, V. Jahnke, H. Jeong et al., \emph{Pole-skipping of scalar and vector fields in hyperbolic space: conformal blocks and holography},  \emph{J. High Energ. Phys.} {\bf 2020} (2020) 111 [arXiv:2006.00974].
\bibitem{Choi}C. Choi, M. Mezei and G. S{\'a}rosi, \emph{Pole skipping away from maximal chaos} [arXiv: 2010.08558].
\bibitem{Karunava}K. Sil, \emph{Pole skipping and chaos in anisotropic plasma: a holographic study} [arXiv:2012.07710].
\bibitem{Yongjun2}Y. Ahn, V. Jahnke, H. Jeong et al., \emph{Classifying pole-skipping points} [arXiv:2010.16166].
\bibitem{Iqbal}N. Iqbal and H. Liu, \emph{Universality of the hydrodynamic limit in AdS/CFT and the membrane paradigm}, \emph{Phys. Rev. D.} {\bf 79} (2009) 025023 [arXiv:0809.2808].
\bibitem{Grozdanov3}S. Grozdanov, P.K. Kovtun, A.O. Starinets et al., \emph{The complex life of hydrodynamic modes}, \emph{J. High Energ. Phys.} {\bf 2019} (2019) 97	[arXiv:1904.12862].
\bibitem{Inkof}G. A. Inkof, J. M. C. Kuppers, J. M. Link et al., Quantum critical scaling and holographic bound for transport coefficients near Lifshitz points [arXiv:1907.05744].
\bibitem{Ge2}X. Ge, Y. Tian, S. Wu et al., \emph{Linear and quadratic in temperature resistivity from holography}, \emph{J. High Energ. Phys.} {\bf 2016} (2016) 128 [arXiv:1606.07905].
\bibitem{Ge1}X. Ge, Y. Tian, S. Wu, and S. Wu, \emph{Hyperscaling violating black hole solutions and magneto-thermoelectric DC conductivities in holography},  \emph{Phys. Rev. D.} {\bf 96} (2017) 046015 [arXiv:1606.05959].
\bibitem{Ge3}X. Ge, S. Sin and S. Wu, \emph{Universality of DC electrical conductivity from holography}, \emph{Physics Letters B.}, {\bf 767} (2017) 63 [arXiv:1512.01917].
\bibitem{Chen}Z. Chen, X. Ge, S. Wu et al., \emph{Magnetothermoelectric DC conductivities from holography models with hyperscaling factor in Lifshitz spacetime}, \emph{Nuclear Physics B.} {\bf 924} (2017) 387 [arXiv:1709.08428].
\bibitem{Gouteraux}B. Gout{\'e}raux, \emph{Charge transport in holography with momentum dissipation}, \emph{J. High Energ. Phys.} {\bf 2014} (2014) 181 [arXiv:1401.5436].
\bibitem{Thomas} T. Faulkner, H. Liu, J. McGreevy and David Vegh, \emph{Emergent quantum criticality, Fermi surfaces, and
AdS$_2$}, \emph{Phys. Rev. D.} {\bf 83} (2011) 125002 [arXiv:0907.2694].
\bibitem{Robert} Robert M. Wald, \emph{General Relativity}, University Of Chicago Press (1984).

\end{thebibliography}
\end{document}